\documentclass[twocolumn]{aastex7}

\usepackage{bmpsize}

\newcommand{\mic}{\ensuremath{\mu\rm m}}

\newcommand{\new}[1]{{\color{red} #1}}
\renewcommand{\new}[1]{#1}

\begin{document}

\title{The JADES Origins Field: A New JWST Deep Field in the JADES Second NIRCam Data Release}

ma\author[0000-0002-2929-3121]{Daniel J.\ Eisenstein} \affiliation{Center for Astrophysics $|$ Harvard \& Smithsonian, 60 Garden St., Cambridge MA 02138 USA}\email{deisenstein@cfa.harvard.edu}
\author[0000-0002-9280-7594]{Benjamin D.\ Johnson} \affiliation{Center for Astrophysics $|$ Harvard \& Smithsonian, 60 Garden St., Cambridge MA 02138 USA}\email{benjamin.johnson@cfa.harvard.edu}
\author[0000-0002-4271-0364]{Brant Robertson} \affiliation{Department of Astronomy and Astrophysics University of California, Santa Cruz, 1156 High Street, Santa Cruz CA 96054 USA}\email{brant@ucsc.edu}
\author[0000-0002-8224-4505]{Sandro Tacchella} \affiliation{Kavli Institute for Cosmology, University of Cambridge, Madingley Road, Cambridge CB3 0HA, UK} \affiliation{Cavendish Laboratory, University of Cambridge, 19 JJ Thomson Avenue, Cambridge CB3 0HE, UK}\email{st578@cam.ac.uk}
\author[0000-0003-4565-8239]{Kevin Hainline} \affiliation{Steward Observatory, University of Arizona, 933 N. Cherry Avenue, Tucson AZ 85721 USA}\email{kevinhainline@arizona.edu}
\author[0000-0002-6780-2441]{Peter Jakobsen} \affiliation{Cosmic Dawn Center (DAWN), Copenhagen, Denmark} \affiliation{Niels Bohr Institute, University of Copenhagen, Jagtvej 128, DK-2200, Copenhagen, Denmark}\email{pjakobsen@nbi.ku.dk}
\author[0000-0002-4985-3819]{Roberto Maiolino} \affiliation{Kavli Institute for Cosmology, University of Cambridge, Madingley Road, Cambridge CB3 0HA, UK} \affiliation{Cavendish Laboratory, University of Cambridge, 19 JJ Thomson Avenue, Cambridge CB3 0HE, UK} \affiliation{Department of Physics and Astronomy, University College London, Gower Street, London WC1E 6BT, UK}\email{rm665@cam.ac.uk}
\author[0000-0001-8470-7094]{Nina Bonaventura} \affiliation{Cosmic Dawn Center (DAWN), Copenhagen, Denmark} \affiliation{Niels Bohr Institute, University of Copenhagen, Jagtvej 128, DK-2200, Copenhagen, Denmark} \affiliation{Steward Observatory, University of Arizona, 933 N. Cherry Avenue, Tucson AZ 85721 USA}\email{nbonaventura@arizona.edu}
\author[0000-0002-8651-9879]{Andrew J.\ Bunker} \affiliation{Department of Physics, University of Oxford, Denys Wilkinson Building, Keble Road, Oxford OX1 3RH, UK}\email{andy.bunker@physics.ox.ac.uk}
\author[0000-0002-0450-7306]{Alex J.\ Cameron} \affiliation{Department of Physics, University of Oxford, Denys Wilkinson Building, Keble Road, Oxford OX1 3RH, UK}\email{alex.cameron@physics.ox.ac.uk}
\author[0000-0002-1617-8917]{Phillip A.\ Cargile}\affiliation{Center for Astrophysics $|$ Harvard \& Smithsonian, 60 Garden St., Cambridge MA 02138 USA}\email{pcargile@cfa.harvard.edu}
\author[0000-0002-9551-0534]{Emma Curtis-Lake} \affiliation{Centre for Astrophysics Research, Department of Physics, Astronomy and Mathematics, University of Hertfordshire, Hatfield AL10 9AB, UK }\email{e.curtis-lake@herts.ac.uk}
\author[0000-0002-8543-761X]{Ryan Hausen} \affiliation{Department of Physics and Astronomy, The Johns Hopkins University, 3400 N. Charles St., Baltimore, MD 21218}\email{rhausen@jhu.edu}
\author[0000-0001-8630-2031]{D\'avid Pusk\'as} \affiliation{Kavli Institute for Cosmology, University of Cambridge, Madingley Road, Cambridge CB3 0HA, UK} \affiliation{Cavendish Laboratory, University of Cambridge, 19 JJ Thomson Avenue, Cambridge CB3 0HE, UK}\email{dp670@cam.ac.uk}
\author[0000-0002-7893-6170]{Marcia Rieke} \affiliation{Steward Observatory, University of Arizona, 933 N. Cherry Avenue, Tucson AZ 85721 USA}\email{mrieke@gmail.com}
\author[0000-0002-4622-6617]{Fengwu Sun} \affiliation{Steward Observatory, University of Arizona, 933 N. Cherry Avenue, Tucson AZ 85721 USA}\affiliation{Center for Astrophysics $|$ Harvard \& Smithsonian, 60 Garden St., Cambridge MA 02138 USA}\email{fengwu.sun@cfa.harvard.edu}
\author[0000-0001-9262-9997]{Christopher N.\ A.\ Willmer} \affiliation{Steward Observatory, University of Arizona, 933 N. Cherry Avenue, Tucson AZ 85721 USA}\email{cnaw@as.arizona.edu}
\author[0000-0002-4201-7367]{Chris Willott} \affiliation{NRC Herzberg, 5071 West Saanich Rd, Victoria, BC V9E 2E7, Canada}\email{chris.willott@nrc.ca}
\author[0000-0002-8909-8782]{Stacey Alberts} \affiliation{Steward Observatory, University of Arizona, 933 N. Cherry Avenue, Tucson AZ 85721 USA}\email{salberts@stsci.edu}
\author[0000-0001-7997-1640]{Santiago Arribas} \affiliation{Centro de Astrobiolog\'ia (CAB), CSIC–INTA, Cra. de Ajalvir Km.~4, 28850- Torrej\'on de Ardoz, Madrid, Spain}\email{arribas@cab.inta-csic.es}
\author[0000-0003-0215-1104]{William M.\ Baker} \affiliation{Kavli Institute for Cosmology, University of Cambridge, Madingley Road, Cambridge CB3 0HA, UK} \affiliation{Cavendish Laboratory, University of Cambridge, 19 JJ Thomson Avenue, Cambridge CB3 0HE, UK}\email{william.baker@nbi.ku.dk}
\author[0000-0002-4735-8224]{Stefi Baum} \affiliation{Department of Physics and Astronomy, University of Manitoba, Winnipeg, MB R3T 2N2, Canada}\email{stefi.baum@umanitoba.ca}
\author[0000-0003-0883-2226]{Rachana Bhatawdekar} \affiliation{ European Space Agency (ESA), European Space Astronomy Centre (ESAC), Camino Bajo del Castillo s/n, 28692 Villanueva de la Ca\~nada, Madrid, Spain} \affiliation{European Space Agency, ESA/ESTEC, Keplerlaan 1, 2201 AZ Noordwijk, NL}\email{rachanab@gmail.com}
\author[0000-0002-6719-380X]{Stefano Carniani} \affiliation{Scuola Normale Superiore, Piazza dei Cavalieri 7, I-56126 Pisa, Italy}\email{stefano.carniani@sns.it}
\author[0000-0003-3458-2275]{Stephane Charlot} \affiliation{Sorbonne Universit\'e, CNRS, UMR 7095, Institut d'Astrophysique de Paris, 98 bis bd Arago, 75014 Paris, France}\email{charlot@iap.fr}
\author[0000-0002-2178-5471]{Zuyi Chen} \affiliation{Steward Observatory, University of Arizona, 933 N. Cherry Avenue, Tucson AZ 85721 USA}\email{zychen@arizona.edu}
\author[0000-0002-7636-0534]{Jacopo Chevallard} \affiliation{Department of Physics, University of Oxford, Denys Wilkinson Building, Keble Road, Oxford OX1 3RH, UK}\email{chevalla@iap.fr}
\author[0000-0002-2678-2560]{Mirko Curti} \affiliation{European Southern Observatory, Karl-Schwarzschild-Strasse 2, 85748 Garching, Germany} \affiliation{Kavli Institute for Cosmology, University of Cambridge, Madingley Road, Cambridge CB3 0HA, UK} \affiliation{Cavendish Laboratory, University of Cambridge, 19 JJ Thomson Avenue, Cambridge CB3 0HE, UK}\email{mirko.curti@eso.org}
\author[0000-0002-4781-9078]{Christa DeCoursey} \affiliation{Steward Observatory, University of Arizona, 933 N. Cherry Avenue, Tucson AZ 85721 USA}\email{cndecoursey@arizona.edu}
\author[0000-0003-2388-8172]{Francesco D'Eugenio} \affiliation{Kavli Institute for Cosmology, University of Cambridge, Madingley Road, Cambridge CB3 0HA, UK} \affiliation{Cavendish Laboratory, University of Cambridge, 19 JJ Thomson Avenue, Cambridge CB3 0HE, UK}\email{fd391@cam.ac.uk}
\author[0000-0002-2380-9801]{Anna de Graaff} \affiliation{Max-Planck-Institut f\"ur Astronomie, K\"onigstuhl 17, D-69117, Heidelberg, Germany}\email{degraaff@mpia.de}
\author[0000-0003-1344-9475]{Eiichi Egami} \affiliation{Steward Observatory, University of Arizona, 933 N. Cherry Avenue, Tucson AZ 85721 USA}\email{egami@arizona.edu}
\author[0000-0003-4337-6211]{Jakob M.\ Helton} \affiliation{Steward Observatory, University of Arizona, 933 N. Cherry Avenue, Tucson AZ 85721 USA}\email{jakobhelton@psu.edu}
\author[0000-0001-7673-2257]{Zhiyuan Ji} \affiliation{Steward Observatory, University of Arizona, 933 N. Cherry Avenue, Tucson AZ 85721 USA}\email{zhiyuanji@arizona.edu}
\author[0000-0002-0267-9024]{Gareth C.\ Jones} \affiliation{Department of Physics, University of Oxford, Denys Wilkinson Building, Keble Road, Oxford OX1 3RH, UK}\email{gj283@cam.ac.uk}
\author[0000-0002-5320-2568]{Nimisha Kumari} \affiliation{AURA for European Space Agency, Space Telescope Science Institute, 3700 San Martin Drive. Baltimore, MD, 21210}\email{kumari@stsci.edu}
\author[0000-0002-4034-0080]{Nora L\"utzgendorf} \affiliation{European Space Agency, Space Telescope Science Institute, Baltimore, Maryland, US}\email{nora.luetzgendorf@esa.int}
\author[0000-0003-4323-0597]{Isaac Laseter} \affiliation{Department of Astronomy, University of Wisconsin-Madison, 475 N. Charter St., Madison, WI 53706 USA}\email{laseter@wisc.edu}
\author[0000-0002-3642-2446]{Tobias J.\ Looser} \affiliation{Kavli Institute for Cosmology, University of Cambridge, Madingley Road, Cambridge CB3 0HA, UK}\affiliation{Cavendish Laboratory, University of Cambridge, 19 JJ Thomson Avenue, Cambridge CB3 0HE, UK}\email{tobias.looser@cfa.harvard.edu}
\author[0000-0002-6221-1829]{Jianwei Lyu} \affiliation{Steward Observatory, University of Arizona, 933 N. Cherry Avenue, Tucson AZ 85721 USA}\email{jianwei@arizona.edu}
\author[0000-0003-0695-4414]{Michael V.\ Maseda} \affiliation{Department of Astronomy, University of Wisconsin-Madison, 475 N. Charter St., Madison, WI 53706 USA}\email{maseda@astro.wisc.edu}
\author[0000-0002-7524-374X]{Erica Nelson} \affiliation{Department for Astrophysical and Planetary Science, University of Colorado, Boulder, CO 80309 USA}\email{Erica.June.Nelson@Colorado.edu}
\author[0000-0002-7392-7814]{Eleonora Parlanti} \affiliation{Scuola Normale Superiore, Piazza dei Cavalieri 7, I-56126 Pisa, Italy}\email{eleonora.parlanti@sns.it}
\author[0000-0003-2662-6821]{Bernard J. Rauscher}\affiliation{Observational Cosmology Laboratory, NASA Goddard Space Flight Center, 8800 Greenbelt Road, Greenbelt MD 20771 USA}\email{Bernard.J.Rauscher@nasa.gov}
\author[0000-0002-7028-5588]{Tim Rawle} \affiliation{European Space Agency (ESA), European Space Astronomy Centre (ESAC), Camino Bajo del Castillo s/n, 28692 Villafranca del Castillo, Madrid, Spain}\email{tim.rawle@esa.int}
\author[0000-0003-2303-6519]{George Rieke} \affiliation{Steward Observatory and Dept of Planetary Sciences, University of Arizona 933 N. Cherry Avenue Tucson AZ 85721 USA}\email{ghrieke@gmail.com}
\author[0000-0003-4996-9069]{Hans-Walter Rix}\affiliation{Max-Planck-Institut f\"ur Astronomie, K\"onigstuhl 17, D-69117, Heidelberg, Germany}\email{rix@mpia.de}
\author[0000-0002-0303-499X]{Wiphu Rujopakarn} \affiliation{National Astronomical Research Institute of Thailand, Don Kaeo, Mae Rim, Chiang Mai 50180, Thailand}\affiliation{Department of Physics, Faculty of Science, Chulalongkorn University, 254 Phayathai Road, Pathumwan, Bangkok 10330, Thailand}\email{wiphu@narit.or.th}
\author[0000-0001-9276-7062]{Lester Sandles} \affiliation{Kavli Institute for Cosmology, University of Cambridge, Madingley Road, Cambridge CB3 0HA, UK} \affiliation{Cavendish Laboratory, University of Cambridge, 19 JJ Thomson Avenue, Cambridge CB3 0HE, UK}\email{ls861@cam.ac.uk}
\author[0000-0001-5333-9970]{Aayush Saxena} \affiliation{Department of Physics, University of Oxford, Denys Wilkinson Building, Keble Road, Oxford OX1 3RH, UK} \affiliation{Department of Physics and Astronomy, University College London, Gower Street, London WC1E 6BT, UK}\email{aayush.saxena@physics.ox.ac.uk}
\author{Jan Scholtz} \affiliation{Kavli Institute for Cosmology, University of Cambridge, Madingley Road, Cambridge CB3 0HA, UK} \affiliation{Cavendish Laboratory, University of Cambridge, 19 JJ Thomson Avenue, Cambridge CB3 0HE, UK}\email{honzascholtz@gmail.com}
\author[0000-0001-8225-8969]{Katherine Sharpe} \affiliation{Center for Astrophysics $|$ Harvard \& Smithsonian, 60 Garden St., Cambridge MA 02138 USA}\affiliation{Department of Astronomy, University of California, Berkeley, 501 Campbell Hall \#3411, Berkeley, CA 94720, USA}\email{kesharpe@berkeley.edu}
\author[0000-0003-4702-7561]{Irene Shivaei} \affiliation{Centro de Astrobiolog\'ia (CAB), CSIC-INTA, Ctra. de Ajalvir km 4, Torrej\'on de Ardoz, E-28850, Madrid, Spain}\email{ishivaei@cab.inta-csic.es}
\author[0000-0003-4770-7516]{Charlotte Simmonds} \affiliation{Kavli Institute for Cosmology, University of Cambridge, Madingley Road, Cambridge CB3 0HA, UK} \affiliation{Cavendish Laboratory, University of Cambridge, 19 JJ Thomson Avenue, Cambridge CB3 0HE, UK}\email{cs2210@cam.ac.uk}
\author[0000-0001-8034-7802]{Renske Smit} \affiliation{Astrophysics Research Institute, Liverpool John Moores University, 146 Brownlow Hill, Liverpool L3 5RF, UK}\email{R.Smit@ljmu.ac.uk}
\author[0000-0001-8426-1141]{Michael W.\ Topping} \affiliation{Steward Observatory, University of Arizona, 933 N. Cherry Avenue, Tucson AZ 85721 USA}\email{michaeltopping@arizona.edu}
\author[0000-0003-4891-0794]{Hannah \"Ubler} \affiliation{Kavli Institute for Cosmology, University of Cambridge, Madingley Road, Cambridge CB3 0HA, UK} \affiliation{Cavendish Laboratory, University of Cambridge, 19 JJ Thomson Avenue, Cambridge CB3 0HE, UK}\affiliation{Max-Planck-Institut f\"ur extraterrestrische Physik, Gie{\ss}enbachstra{\ss}e 1, 85748 Garching, Germany}\email{hannah@mpe.mpg.de}
\author[0000-0001-8349-3055]{Giacomo Venturi} \affiliation{Scuola Normale Superiore, Piazza dei Cavalieri 7, I-56126 Pisa, Italy}\email{giacomo.venturi1@sns.it}
\author[0000-0003-2919-7495]{Christina C.\ Williams} \affiliation{NSF’s National Optical-Infrared Astronomy Research Laboratory, 950 North Cherry Avenue, Tucson, AZ 85719 USA}\email{christina.williams@noirlab.edu}
\author[0000-0002-7595-121X]{Joris Witstok} \affiliation{Kavli Institute for Cosmology, University of Cambridge, Madingley Road, Cambridge CB3 0HA, UK} \affiliation{Cavendish Laboratory, University of Cambridge, 19 JJ Thomson Avenue, Cambridge CB3 0HE, UK}\email{joris.witstok@nbi.ku.dk}
\author[0000-0001-5962-7260]{Charity Woodrum} \affiliation{Steward Observatory, University of Arizona, 933 N. Cherry Avenue, Tucson AZ 85721 USA}\email{cwoodrum@arizona.edu}

\begin{abstract}
We summarize the properties and initial data release of the JADES Origins Field (JOF),
the longest single pointing yet imaged
with the James Webb Space Telescope (JWST).  
This field falls within the GOODS-S region about 8' south-west of the Hubble Ultra Deep Field (HUDF), where it was formed initially in Cycle 1 as a parallel field of HUDF spectroscopic observations within the JWST Advanced
Deep Extragalactic Survey (JADES).  
This imaging was greatly extended in Cycle 2 program 3215,
which observed the JOF for 5 days in six medium-band
filters, seeking robust candidates for $z>15$ galaxies. 
This program also includes ultra-deep parallel NIRSpec spectroscopy (up to 91 hours on-source, summing over the dispersion modes) on the HUDF.
Cycle 3 observations from program 4540 added 20 hours of NIRCam slitless spectroscopy and F070W imaging to the JOF.
With these three campaigns, the JOF was observed for 380 open-shutter hours with NIRCam using 15 imaging filters and 2 grism bandpasses.  Further, parts of the JOF have deep 43 hr MIRI observations in F770W.
Taken together, the JOF is one of the most compelling deep fields available
with JWST and a powerful window into the early Universe.  This paper presents the second data release from JADES, featuring the imaging and catalogs from the year 1 JOF observations.
\end{abstract}

\keywords{early universe — galaxies: evolution — galaxies: high-redshift}

\section{Introduction}

A primary goal of the James Webb Space Telescope (JWST) is to study galaxies to the earliest observable epochs, when the first massive assemblies of gas and stars set in motion the long play of galaxy evolution. Precisely timing when the first galaxies formed after the Big Bang is a key goal of observational cosmology and one of the last great unknowns in our history of the Universe: When and how did the first galaxies form? The primordial star formation and the formation of galaxies in the first 300 Myr (redshift $z\gtrsim 15$) after the Big Bang is a sensitive probe of the physics governing the conversion of gas into stars at early times and different dark matter models \citep[e.g.,][]{Behroozi15, Dayal18, Boylan-kolchin22, Kannan22}. Additionally, these first, unexplored galaxies might harbor surprises such as metal-free stellar populations (Population III) or rapidly accreting young black hole seeds, thereby shedding light onto possible pathways to the formation of the billion-solar-mass black holes observed at $z\sim6$ \citep[e.g.,][]{Bromm11, Xu13,Smith19,Vikaeus22,Volonteri21,Trinca23,Schneider23}.
Deep fields, both in imaging and associated spectroscopy, are a key opportunity for this work, allowing the astronomy community to coordinate its multi-wavelength resources to best advantage.

Cycle 1 observations with JWST succeeded quickly in pushing the redshift frontier beyond what had been seen with HST.  Relying on the complete scattering of photons shortward of Ly$\alpha$ 1216\AA\ by the intergalactic medium (IGM), many F115W dropouts at $z\sim 10$ and even some F150W dropouts at $z\sim 13$ were detected as photometric candidates in numerous fields \citep[e.g.,][]{Castellano22,Finkelstein22,Finkelstein24,Naidu22,  Adams23,Austin23,Bouwens22,Donnan23,Harikane22, PerezGonzalez23,Robertson22, Whitler23,Yan23, Atek23,Franco23, Hainline23}.  In the first year, numerous galaxies at $z>10$, including two at $z\sim 13$, were spectroscopically confirmed with long NIRSpec exposures \citep{Curtis-lake22, Heintz22, Morishita22,Roberts-borsani22,ArrabalHaro23b,Bunker23gnz11,Fujimoto23,Sanders23,Tang23,Wang2023a,Casey24,Hsiao23,McLeod24}, along with a larger number of galaxies out to $z\sim 9.5$ where the much brighter [OIII] and H$\beta$ lines are available in the NIRSpec bands.

These early results demonstrated not only JWST's unrivalled capabilities in successfully exploring the early Universe, but also its potential for pushing the redshift frontier even further.
Indeed, it is evident that JWST---with its excellent design and on-orbit performance \citep{Rigby22}---is fully capable of probing back to even earlier cosmic times. Within this context, we note that the Cycle 1 programs were designed to go deep in F115W and F150W, not in F200W.
The few hints of any F200W dropouts are uncertain, potentially mimicked by $z\approx4$--5 galaxies with dust-reddened continua and well-placed strong rest-optical emission lines \citep{Fujimoto22, Naidu22_schrodinger, Zavala22,Arrabalharo23,PerezGonzalez23b}.

We describe here the construction of a new deep field, the JADES Origins Field (JOF), designed to push our exploration of early galaxies further than ever before possible.  This field was initiated by the JWST Advanced Deep Extragalactic Survey \citep[JADES; ][]{Eisenstein23}  as 
JWST NIRCam coordinated parallel observations to a NIRSpec multi-object spectroscopy (MOS) pointing focused on the Hubble
Ultra Deep Field \citep[HUDF; ][]{Beckwith06} within the GOODS-S field \citep{Giavalisco04}.  
These Cycle 1 observations of JADES were described 
in \citet{Eisenstein23} and \citet{Bunker23}.  
New data in Cycle 2 and 3 brought this
field to nearly 10 mission days of NIRCam imaging and slitless spectroscopy
in addition to over 7 mission days of deep MIRI imaging, making it the deepest JWST imaging yet achieved.  The geometry of this field is well suited for
future JWST work, as the parallels can span between the JADES Deep mosaic around the HUDF and the JOF at observing windows that have low zodiacal background.

In this paper, we detail the scientific design and implementation of the
Cycles 2 and 3 observations.  
Relative to other deep fields with JWST, this combined data set features very deep
medium-band imaging in a total of 15 NIRCam filters.  The medium-band
imaging is motivated by the goal of having finer sampling of the spectral
energy distribution for searches for Ly$\alpha$ dropouts at $z>15$.  
As we argue in \S~\ref{sec:cycle2} (and as mentioned above), broad-band NIRCam data alone is 
susceptible to selection of Balmer-break and emission-line interlopers 
from mid-redshift, and this failure mode appears to become more pernicious
at $z>15$, because of the rarity of these galaxies and because of the
more limited number of applicable filters.  Using medium-bands F162M, 
F182M, F210M, F250M, F300M, and F335M can exclude mid-redshift
emission lines and allow for detection of more distinctive breaks.
These data provide robust observational constraints on the number density of $z>15$ galaxies and their rest-UV luminosities, colors, and sizes \citep{Robertson24}

Of course, these exceptionally deep images, up to 46 hours per 
filter and covering about 9 square arcminutes, are of high value
for the study of intermediate-redshift galaxies as well.  
JADES has conducted 5 NIRSpec MOS pointings on the JOF itself; these data will be presented in JADES Data Release 4 (Curtis-Lake et al., in prep; Scholtz et al., in prep.).  An additional pointing using a novel ultra-high multiplex layout for the G235M and G395M grating was taken as part of program 3215 and will be presented in D'Eugenio et al.\ (in prep.).

To augment this slit spectroscopy, the NIRCam instrument team conducted NIRCam slitless spectroscopy in Cycle 3 
in the F322W2, F356W, and F444W filters, reaching up to 9 hr of on-source integration time.
These will provide spectroscopic redshifts for hundreds of galaxies at $3.7<z<9$ with the strong H$\alpha$ and/or 
[OIII] 5007\AA\ lines.

A further opportunity of the Cycle 1 and 2 data is the observation of a very deep NIRSpec MOS pointing on the UDF.
The Cycle 1 data was already released in \citet{Bunker23}; here we give an overview of the Cycle 2 pointing, which targeted faint high-redshift galaxies with exposure times up to 91 hrs on-source.  Further description of the NIRSpec target selection and data properties is presented in JADES Data Release 3 \citep{DEugenio25}.

In addition to an overview of the upcoming programs, this paper presents the initial release of NIRCam imaging data and catalogs on this field, already some of the deepest imaging obtained by JWST.  The release includes medium-depth flanking fields that connect this footprint to the HUDF region released in the JADES first data release \citep{Rieke23}.  Reductions of the deep imaging results will be in an upcoming data release.

In \S~2--4, we present the Cycle 1, 2, and 3 programs in the JOF.  \S~5 then presents the initial data release, and we summarize in \S~6.

\section{Cycle 1: Beginning this field as a Coordinated Parallel}

The imaging in this field began as the NIRCam Coordinated Parallel of 
program 1210 (PI: L\"utzgendorf), which executed the JADES Deep/HST
NIRSpec micro-shutter array (MSA) observations in and around the HUDF \citep{Bunker23}.
Fortuitously, the position angle of these observations put these NIRCam
parallels on top of further data from program 1180 (PI: Eisenstein) that
included medium-depth NIRCam imaging and deep MIRI F770W imaging, the
latter resulting from coordinated parallels of deep NIRCam imaging around
the HUDF.  
These observations are described in \citet{Eisenstein23}.  In brief, 
the 1210 parallel provides 55 open-shutter dual-filter hours of NIRCam
imaging in 9 filters: F090W, F115W, F150W, F200W, F277W, F335M, F356W, F410M, 
and F444W.  The 1180 medium imaging overlaps and extends this geometry at flanking depth
in all filters save F335M.  The MIRI imaging reach 43 hr depth in
F770W in each of 4 adjoining fields.

Further, an assigned observing window in Year 2 allowed us to place one Coordinated 
Parallel NIRCam field from the JADES Medium/JWST tier (program 1286, PI: L\"utzgendorf)
in a partially overlapping location about 1$'$ south of the 1210 pointing.  This is 11 hours
open-shutter of additional data in all the previous nine filters plus F070W. It adds further depth to the JOF imaging, but also makes a deeper region (being the superposition of two medium surveys) that is more square and hence better suited for the NIRSpec MOS footprint.  The extra NIRCam imaging also overlaps more of the deep MIRI data.

\begin{figure}[t]
\noindent
\includegraphics[width=\columnwidth]{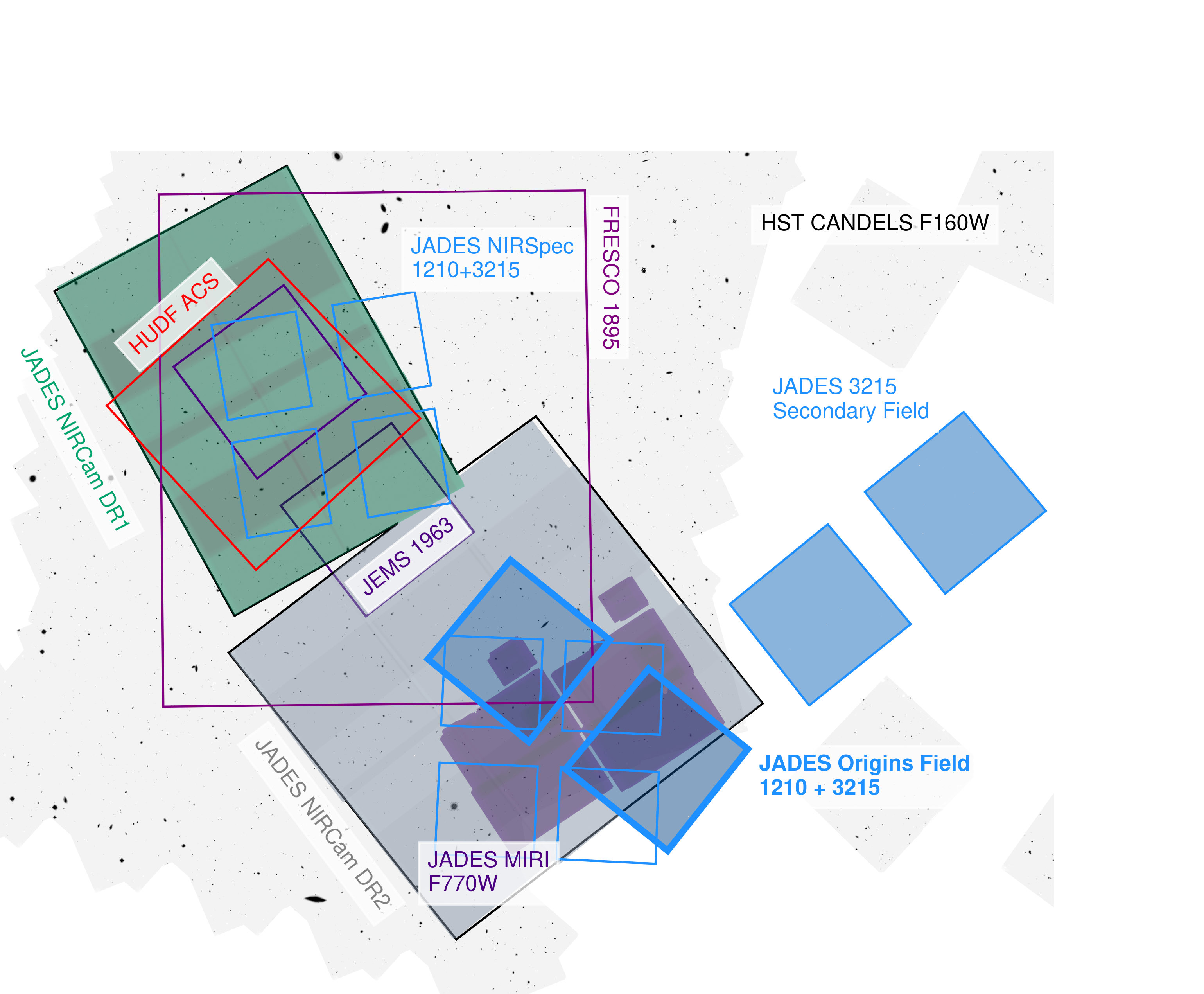}
\caption{\label{fig:layout}%
The layout of data sets in the GOODS-S field most immediate to this paper, overlaid on a background F160W image from HST CANDELS \citep{Whitaker19}.  JADES DR1 and DR2 are shown in green and grey, using the F356W exposure map. The parallel imaging in 1210 is the deepest portion; program 3215 extends this with 6 medium-bands.  The 3215 secondary field location lies to the northwest; it overlaps other JADES NIRCam imaging from program 1286 (not shown).  The NIRSpec MSA footprints for the two 3215 pointings are shown.
We also display the footprints of the HUDF ACS field \citep{Beckwith06}, 
FRESCO grism \citep[][program 1895;]{Oesch23}, 
and JEMS medium-band \citep[program 1963;][]{Williams2023}, 
as these are immediately supportive of the target selection for the 3215 NIRSpec observations.  There are many other powerful data sets in this region, not shown for brevity!
}
\end{figure}

A simplified view of the GOODS-S region is provided in Figure \ref{fig:layout}, showing the location of the JOF relative to the HUDF, CANDELS, and other JWST programs.

\section{Cycle 2: Deep Medium-Band Imaging with Program 3215}\label{sec:cycle2}

\subsection{Science Motivation: Key Opportunities at \texorpdfstring{$z>15$}{z>15}}

When, how, and why the first stars and black holes formed after the Big Bang is one of the provocative mysteries of the Universe. The galaxies hosting those stars and black holes have remained out of reach until now.  With the advent of JWST, we now have a telescope in space that can directly probe and characterize the formation of those first galaxies. This is a key goal of JWST and observational cosmology because the first galaxies probe unknown physics related to star formation (e.g., the efficiency of how gas is converted into stars in early dark matter halos e.g., \citealt{Behroozi15, Gnedin16, Vogelsberger20, Kannan22, Wilkins23}) and early structure formation \citep[e.g.,][]{Dayal18, Khimey21, Lovell23, Boylan-kolchin22}. While the low metallicity in this first generation of galaxies should slow H$_2$ formation and decrease star-formation efficiency \citep[e.g.,][]{Krumholz12}, the gas-rich, compact nature of early galaxies may result in increased gas surface densities and star-formation efficiencies \citep[e.g.,][]{Krumholz09, Ostriker11, Somerville15}.

There are a wide variety of pre-flight theoretical predictions for the redshift evolution of the UV luminosity function of galaxies
\citep[e.g.,][]{Behroozi15,Mason15,Tacchella18,Kannan22,Wilkins23,Katz21}
These different models typically lead to similar outcomes at $z<8$ regarding number counts and the cosmic star-formation rate density, but they diverge more and more toward higher redshifts due to different treatments of cooling, star formation efficiency, and feedback.
Similarly, predictions for the formation and evolution of supermassive black holes vary widely \citep{Habouzit22}.
With JWST, we have the opportunity to push to the redshift frontier of $z>15$ where the physics of these extreme settings cause large differences, factors of 10 and more.

Of course, the redshift frontier offers the opportunity for surprises and more exotic discoveries.
As dark matter halos are rapidly growing in this early epoch, galaxies at $z>15$ are a sensitive probe of the interplay of gravitational collapse, gas cooling, and energetic feedback.  Indeed, this epoch could even reveal surprises in structure formation and the nature of dark matter: the suppression of small-scale structure in some dark matter models leads to a delayed and subsequently more rapid stellar assembly relative to cold dark matter models (e.g., Gandolfi+ 2022).
The exploration of $z>15$ is the most direct way to constrain the onset of cosmic dawn.
Together with the constraints on the star-formation efficiency and feedback processes at $z>15$, JWST can bridge the gap to ongoing 21cm experiments---such as EDGES \citep{Bowman18}, SARAS 3 \citep{Singh2022} and REACH \citep{DeLeraAcedo22}---and thereby creating new synergies to constrain the formation of the first stars and galaxies.

\begin{figure*}[t!]
\noindent\includegraphics[width=\textwidth]{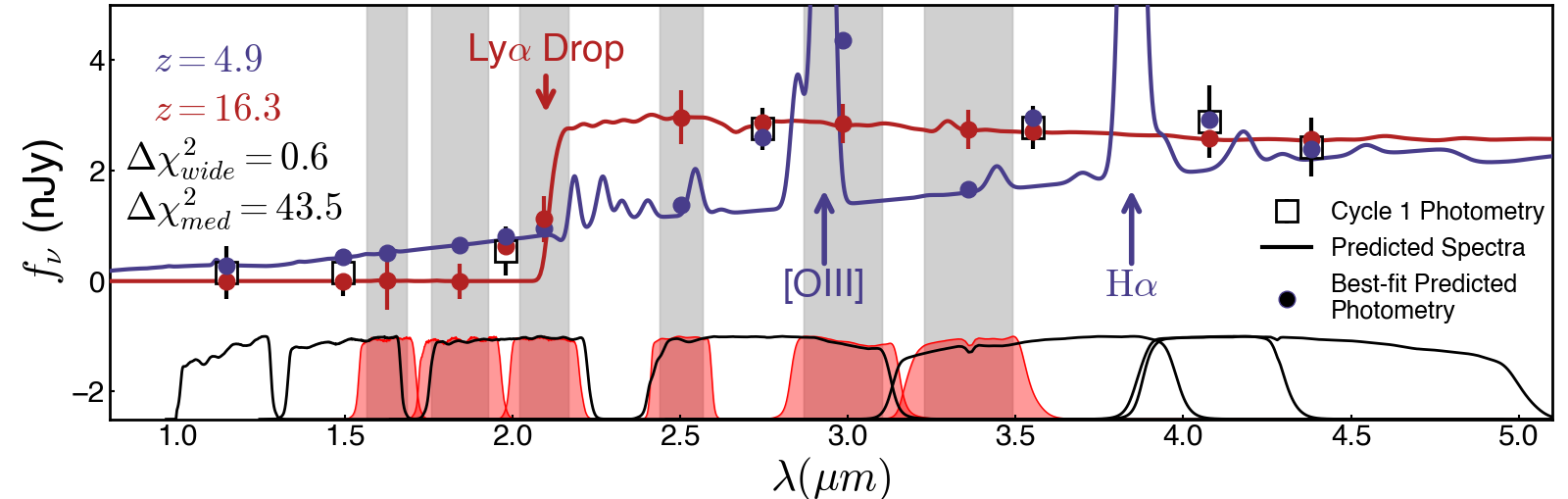}
\caption{\label{fig:mb}%
A demonstration of the ability of medium-bands to starkly differentiate between $z>15$ galaxies and $z\sim5$ interlopers.
Starting from the photometry of the CEERS $z\approx16$ candidate \citep{Donnan23} scaled fainter to $S/N=7$ in F277W and F356W in our survey field, we perform fits to the Cycle 1 wide-band photometry, first for $z\sim16$ solutions and then for $z\sim5$ solutions, using the Prospector galaxy spectrum modeling code \citep{Johnson21}. With the Cycle 1 bands alone, one can get acceptable fits at either redshift ($\Delta\chi^2=0.6$); these best-fit smoothed spectra are plotted.  We then predict the medium-band photometry from each model.  These results clearly distinguish the two hypotheses: the F182M and F210M bands indicate a well-localized break, while the F250M and F300M bands completely exclude the emission lines that the $z\sim5$ solution requires.  Our photometric depth yields $\Delta\chi^2=43$, an overwhelming confirmation.
}\end{figure*}

In addition to delivering number density constraints at $z>15$, JWST observations constrain the properties of these first galaxies. Specifically, these data are sensitive to the total rest-frame UV luminosity, UV continuum slope $\beta$, and the morphology of those first structures. The UV continuum slope is a well-known diagnostic for the stellar populations in galaxies \citep{Bouwens09, Wilkins11, Finkelstein12, Bhatawdekar21}, primarily affected by the dust attenuation and metallicity.
Together with the UV luminosity, one can derive dust-corrected star-formation rates (SFRs) and infer the cosmic SFR density at $z>15$. Furthermore, an estimate of $\beta$ probes the chemical enrichment and dust physics in the earliest galaxies, extending studies at $z\sim7-10$ from lower redshifts \citep[e.g.,][]{Ferrara22}. Very blue UV slopes are a potential signposts of Population III stars or accreting black holes \citep{Schaerer03, Dayal12}. 

The structure and morphology of galaxies is a well-known tracer of galaxy assembly \citep[e.g.,][]{Conselice14}.  JWST images reveal marginally-resolved galaxies at $z\approx10-13$ with compact ($<500$ pc) sizes \citep{Robertson22, Ono22}. Deep imaging at $z>15$ can deliver rest-UV size constraints down to a resolution of $\sim$300 pc.  Measurements of the size and shape (even upper limits) are a sensitive probe of stellar and black hole feedback in those galaxies \citep[e.g.,][]{Pillepich19, Wu20}  and could potentially probe the filamentary structure of the gas accretion \citep[e.g.,][]{Pandya19, Ceverino22}.  Further, one can be sensitive to the multiplicity and close-pair fraction of galaxies at $z>15$, probing the importance of mergers and star-formation in filaments in these early galaxies \citep[e.g.,][]{Wang18, Oleary21}.

\subsection{Observational Challenges and a Method to Reach \texorpdfstring{$z>15$}{z>15}}\label{sec:challenges}

\begin{figure}[t!]
\includegraphics[width=\columnwidth]{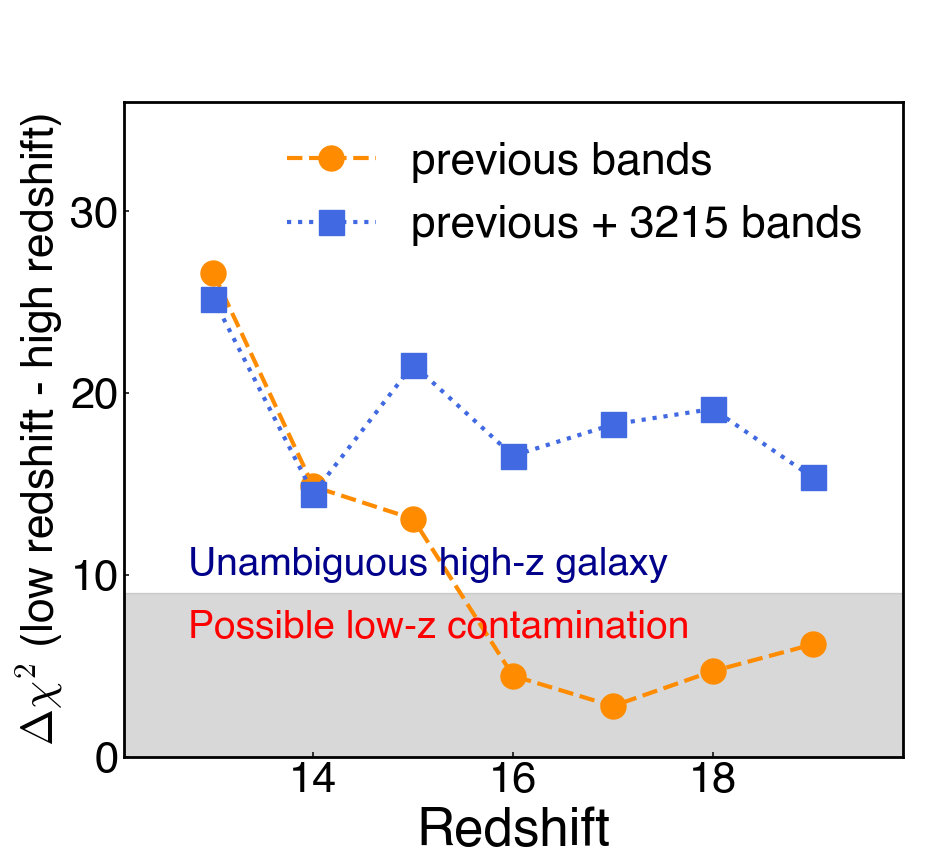}
\caption{\label{fig:mb2}%
We compare two filter strategies in their ability to reject a low-redshift hypothesis for a high-redshift galaxy.  
The first strategy is the Cycle 1 strategy of program 1210, focused on wide bands.  The second strategy is the wide+medium-band strategy adopted with programs 1210 \& 3215.  We have rescaled the errors from the first strategy so as to match the total exposure time of the second, i.e., 2.68 times what program 1210 actually provides. In both, we take the high-redshift fit from Figure \ref{fig:mb} and shift it in redshift, holding the F277W signal-to-noise ratio constant (at 7, for the wide+medium strategy).  At each true redshift, we fit with Prospector and report the $\Delta\chi^2$ between the best low-redshift false solution and the best high-redshift true solution.  This quantifies the rejection of the low-redshift hypothesis.  We find that at $z>15$, the wide+medium strategy provides a clear advantage, nearly 4 times better $\Delta\chi^2$ at fixed exposure time compared to strategy of a typical wide-band survey. 
}
\end{figure}

There are important reasons why pushing for F200W ($z\sim17$) dropouts is substantially harder than F115W ($z\sim10)$ or F150W ($z\sim13$) dropouts:
1) Such galaxies are fainter and rarer.
2) The frequency of extreme emission-line galaxies increases with redshift, and JWST clearly has shown a substantial population of them at $z>3$, where they can create spectral energy distributions (SEDs) that differ from those of purely stellar populations, boosting the rest-optical and enhancing the apparent Balmer drop.
3) The dichroic gap in NIRCam leaves a wider spacing between F200W and F277W that makes it harder to exclude a mildly reddened Balmer-drop galaxy at $z\sim5$, particularly with [OIII] potentially enhancing F277W.
4) Many Cycle 1 programs include only F277W, F356W, F410M, and F444W.\footnote{Notable exceptions are JEMS \citep{Williams2023} and CANUCS \citep{Withers23}.} 
However, the filters beyond 4 microns have reduced sensitivity due to increased zodiacal foreground, so faint galaxies may only be well-detected in the bluer two bands, which in turn can be filled with [OIII] and H$\alpha$.
These higher-redshift break candidates are less compelling than lower redshift examples that have 3 or 4 sensitive bands redward of the proposed break.

Whereas Cycle 1 programs pushed deep in F115W and/or F150W, in program 3215 we adopt a strategy optimized to identify candidates at $z>15$.  In particular, this search must 1) push very deep in the 2 micron region;
2) increase the spectral resolution to isolate the dropout; and 3) include additional filters in the 2.5--4 micron region to eliminate contamination from mid-redshift emission-line galaxies.
We expect that this combination of high-SNR detections, well-sampled dropout features, and systematic control over contaminants can reach the high standard of proof that claims of $z\sim15$ galaxies will require.
The new imaging achieves these requirements by combining the usual NIRCam wide-band filters with well-chosen medium-band filters.  By observing very deep in F182M, F210M, and F250M, the photometry can isolate a sharp break within a narrower and hence more distinctive region.  Further, using the combination of medium and wide-bands in the long-wave (LW) channels reveals strong emission lines that could supply false positive Balmer breaks.

We illustrate this strategy by considering the $z=16$ candidate in the CEERS field \citep{Donnan23}, scaled down by a factor of 30 in flux and assuming the noise levels of  programs 1210 \& 3215.  While now proved to be at $z=4.9$ \citep{Arrabalharo23}, this object was a plausible $z=16$ candidate, showing a drop in F200W by a factor of 4 relative to F277W, no detection in F150W, and a blue continuum in the LW filters.  However, this photometry was also a match to models at $z\approx 5$ with strong optical emission lines and reddening suppression in the UV \citep{Naidu22_schrodinger}.
Figure \ref{fig:mb} shows how these two hypotheses are starkly differentiated by the NIRCam medium bands: the Lyman $\alpha$ break is much steeper than a dust-enhanced Balmer break, and the medium bands easily distinguish continuum from emission lines.

\new{The fits shown in Figure \ref{fig:mb} to the scaled Cycle 1 wide-band photometry use the Prospector galaxy spectrum modeling code \citep{Johnson21}.  The Prospector models use FSPS \citep{fsps} with the MIST isochrones \citep{MIST}, the MILES stellar spectral library \citep{MILES}, and nebular emission following \citet{Byler17}. The fitted model includes a five-bin flexible SFH with broad priors on the ratio of SFR in adjacent bins following \citet{Leja19, Tacchella22} and normalization of a \citet{Kriek13} dust attenuation curve. For the nebular emission, we adopt uniform priors on the log of the nebular ionization parameter and log of the gas-phase metallicity, which is allowed to vary independently of the stellar metallicity.  Attenuation by the IGM follows \citet{Madau95}.  The resulting model space can produce a wide range of galaxy SEDs, making it a compelling test of the ability to reject low-redshift interpretations.  We fit within several relatively narrow redshift ranges and tune the number of live points used for nested sampling with \texttt{dynesty} \citep{Speagle20, Koposov22} to help identify narrow local minima in redshift, which can otherwise be missed \citep[e.g.][]{Naidu22_schrodinger}.}

Shifting the true galaxy in redshift and repeating the fits, Figure \ref{fig:mb2} demonstrates that this suite of filters provides very effective recovery of candidates at $z$ between 15 and 20.
\new{Here we show that for these reference galaxies, a strategy with medium bands produces a much larger $\Delta\chi^2$ rejection of the low-redshift hypothesis.  This plot uses a fixed normalization of F277W signal-to-noise ratio, but brighter or fainter galaxies would simply scale $\Delta\chi^2$ as the square of the flux.  What is important is the ratio of the $\Delta\chi^2$ between the two strategies, as this is proportional to how one would have the scale the exposure times in the background-noise limited case.  One sees that at $z\sim17$, this ratio is about 4 times higher, indicating that the medium-band strategy is more performant.
}

\new{We note that these galaxy models are based on known stellar populations and do not include the extreme Balmer-break phenomenon associated to black-hole stars \citep{Naidu25bh,deGraaff25}, plausibly connected to Little Red Dots \citep{Matthee23}.  Stronger Balmer breaks of course are harder to distinguish from high-redshift candidates, although medium-band photometry will help to distinguish a very abrupt Lyman-$\alpha$ break from the smoother decline observed redward of Balmer break in these AGN \citep{Naidu25bh,deGraaff25}.
Such objects are rare and we don't yet know about their abundance at faint flux levels.  Importantly, they have strong rest-optical lines that should be readily measurable in spectra.  
}

\subsection{Filters, Exposure Times, and Depths}

\newcommand{\nod}{---}
\begin{table*}[t]
\begin{center}
\begin{tabular}{|c|cc|c|cc|c|}
\hline
Filter & \multicolumn{2}{|c|}{Cycle 2, Program 3215}
& Cycle 1 Time & Total Time & 5-$\sigma$ PS & Ly$\alpha$ \\
& $N_{\rm exp}$ & Time (ks) & (ks) & (ks)  & (nJy) & redshift \\
\hline
     F090W  &\nod&  \nod & 55.2 &       55.2 & 3.13 & $5.5<z<7.3$ \\
     F115W  &\nod&  \nod & 72.9 &       72.9 & 2.36 & $7.3<z<9.5$  \\
     F150W  &\nod&  \nod & 55.2 &       55.2 & 2.15 & $9.9<z<12.7$  \\
{\bf F162M} & 30 &  82.5 & \nod &{\bf  82.5} & 2.80 & $11.7<z<13.1$  \\
{\bf F182M} & 60 & 165.0 & \nod &{\bf 165.0} & 1.59 & $13.2<z<15.2$  \\
     F200W  &\nod&  \nod & 38.7 &       38.7 & 2.24 & $13.4<z<17.3$  \\
{\bf F210M} & 45 & 123.8 & \nod &{\bf 123.8} & 2.13 & $15.4<z<17.1$  \\
\hline
{\bf F250M} & 60 & 165.0 & \nod &{\bf 165.0} & 2.29 & $18.8<z<20.3$  \\
     F277W  &\nod&  \nod & 47.0 &       47.0 & 1.86 & $18.9<z<24.8$  \\
{\bf F300M} & 45 & 123.8 & \nod &{\bf 123.8} & 1.80 & $22.3<z<25$  \\
{\bf F335M} & 30 &  82.5 & 24.8 &{\bf 107.3} & 1.84 & $25<z<28$\\
     F356W  &\nod&  \nod & 38.7 &       38.7 & 2.05 & $25<z<32$\\
     F410M  &\nod&  \nod & 55.2 &       55.2 & 3.56 & $31<z<34$ \\
     F444W  &\nod&  \nod & 56.4 &       56.4 & 2.95 & $31<z<40$ \\
\hline
\end{tabular}\hspace*{0.7in}
\end{center}
\vspace*{-12pt}
\caption{\label{tab:exptime}Exposure times per filter.  $N$ is the number of distinct 2750 s exposures in program 3215; these are in groups of 15 because of the 5-visit structure of the NIRSpec observations.  Cycle 1 Time reports the sum of programs 1210 and 1180 (dominated by 1210), and Total sums this with program 3215.  Additional medium-depth imaging provided by parallel imaging from program 1286 is not included because it covers only a portion of the deep field.   We then use the JWST ETC to compute the 5-$\sigma$ depth in $0.2''$ diameter apertures for a background-limited point source; \new{in practice, we have found this to be reasonably close to what we measure from the variance of apertures on blank sky}.  Also listed is the redshift range for which Ly$\alpha$ is in the filter.  We remind that robustly isolating a dropout doesn't come from the Ly$\alpha$ filter, but rather from  measuring the continuum flux from filter(s) just redder and measuring no flux in filter(s) just bluer, so as to establish a sharp break. 
}
\end{table*}

Because of the opportunity to re-use the deep imaging of program 1210, we designed program 3215 to return to the same footprint, matching the position angle exactly.  The imaging exposure times and depths, in both new and old filters, are listed in Table \ref{tab:exptime}, along with the Ly$\alpha$ redshift range for each.

The driving requirement is to go deep in F182M and F210M, as potential dropout bands.  In total, F182M, F200W, and F210M  reached 91 hrs of exposure time, far more than the 11 hrs that 1210 \& 1180 provide and much deeper than any other Cycle 1 program (to our knowledge, program 2079 being deepest with 17 hrs in F200W).
We split the time roughly evenly between these two medium filters, as we do not know at what redshift a candidate might appear.

For cases around $z=15$, F182M may only be a partial dropout, and we must demonstrate a lack of flux shortward of 1.7 $\mu$m.  The existing F150W data would do this, but not as deeply as this 2 $\mu$m data warrants.  We therefore add depth in F162M.  Because the false positive interlopers are mid-$z$ galaxies with steeply falling SEDs, it is more efficient to test the dropout in a narrower band at the red edge of F150W: this includes all of the hypothesized signal and less background noise.  This same argument is why it is better to observe F182M \& F210M, rather than F182M \& F200W: once one has found no flux in F182M, it is more efficient to seek flux in the redder half of F200W with the medium-band filter.

For the matching time in the LW arm, the key filter is F250M, through which one is trying to determine the continuum just longward of a $15<z<19$ Ly$\alpha$ drop {\it or} detect a yet-higher redshift dropout in F250M itself.  But one must guard against mid-redshift emission-line contamination in F250M, and for this we observe F300M and F335M.  F300M is just short enough that if [OIII] 5007 is in F250M, then H$\alpha$ will miss F300M and be found in F335M.  This pairing therefore gives a robust measurement of the continuum.

By placing these observations on the parallel field created by program 1210, these new data build on the 55 hours of imaging depth already invested here.
Given the steep number counts of galaxies at the redshift frontier, it is optimal to maximize the detection rate by integrating deeply on a single NIRCam pointing.
The region in and around the HUDF in GOODS-S has the deepest and most comprehensive imaging and spectroscopy on the sky.  While $z>15$ galaxies could be found anywhere, we site this program in GOODS-S to utilize and build upon the amazing legacy value of this field, while utilizing a position angle that allows ultra-deep MSA spectroscopy on the HUDF at a time of low zodiacal background. 
Of course, these medium-band images also support a wide range of other science cases for intermediate-redshift galaxies, notably emission-line morphologies and SED modeling around the Balmer break, as well as time-domain opportunities from the one-year cadence of the observations.

\new{The results from this high-redshift search are summarized in \S \ref{sec:results}, but we continue next with other aspects of the implementation of program 3215.}

\subsection{Coordinated Parallels with NIRSpec}

\begin{table*}[tb]
\begin{center}
\begin{tabular}{|l|l|ccc|ccccc|}
\hline
& & & & & \multicolumn{5}{c|}{Exposure Times (ksec)} \\
Subsurvey & Program & \# Fields & Subpointings & \# Targets & Prism & G140M & G235M & G395M & G395H \\
\hline
GOODS-S Deep/HST    & 1210 & 1 & 3 & 250 & 100 & 25  & 25  & 25  & 25  \\
GOODS-S Year 2      & 3215 & 1 & 5 & 228 & 134  & 34  & \nodata  & 160  & \nodata  \\
\hline
\end{tabular}
\end{center}
\caption{\label{tab:ns}Summary of the NIRSpec MOS Observations from program 3215, with comparison to those of 1210 at the same location but targetted largely from pre-JWST imaging.
For each program, we list the number of separate MSA fields, as well as the exposure time per disperser in
kiloseconds.  Each field consists of several sub-pointings, each with two nearly identical MSA designs: one for the 
prism and a second for the grating; the latter closes a few shutters to protect certain high-priority spectra from overlap.
The quoted times are summed over the sub-pointings, but not all targets can be placed on all sub-pointings. The number of unique targets in each subsurvey is listed. Note that program 3215 has fewer unique targets than 1210 as in the former the focus was to re-observe as many times as possible each target in the five pointings, so to maximise the exposure time per target.  For 3215, targets on the 5th subpointing will only have Prism data.
The long-pass filter choices for the gratings are F070LP, F170LP, F290LP, and F290LP, respectively.
}
\end{table*}

Program 3215 follows the path of 1210, using the NIRSpec MSA as the prime
instrument and NIRCam as the coordinated parallel.  This results in exceptionally deep spectroscopy on the HUDF and JADES Deep mosaic, about 91 hrs of total
exposure, following on the 55 hrs of program 1210 spectroscopy released in \citet{Bunker23}.

Following the program goal of exploring the high-redshift universe, we optimize
this time toward the study of high-redshift galaxies.  Unlike program 1210, which was
largely targeted from HST ACS optical and WFC3 infrared imaging \citep{Bunker23} save for 
four early high-redshift candidates from JADES \citep{Robertson22,Curtis-lake22}, program 3215  
utilizes NIRCam-selected targets.  This allows a great expansion in the 
number of $z>7$ targets, as the JADES data is comfortably deeper than even the deepest WFC3 HUDF data and much deeper than the flanking CANDELS imaging.

Bringing ultra-deep spectroscopy to these faint high-redshift galaxies addresses many windows to understanding the physical processes associated with the early galaxy formation.
We split the spectroscopic time across three dispersing modes---up to 134 ks on source with G395M (far deeper than program 1210), 160 ks with the Prism, and 34 ks with G140M---to serve a variety of key science drivers in studies of these galaxies.

The choice of obtaining ultra-deep spectra in G395M band comes from the requirement of probing the primary
rest-optical lines at $5<z<9.5$, with H$\alpha$ down to $z=3.5$. 
For galaxies at $z<9.5$, the strong [OIII] 5007,4957\AA\ doublet and 
H$\beta$ line are available to NIRSpec.  H$\alpha$ is available at $z<6.8$.
As has been seen in JADES
and other surveys, the medium-resolution grating (G395M/F290LP) is 
excellent for studying these lines, producing precise redshifts, splitting
closer pairs, and revealing kinematic signatures and broad AGN Balmer lines.

The deep Prism spectra are needed to determine the redshift for faint
galaxies at $z>9.5$ through the detection of
the Ly$\alpha$ drop.  The prism 
also can reveal rest-UV lines if they are strong enough, and generally
provides a powerful route to measure the line flux of isolated lines over
the full NIRSpec wavelength range.

Finally, we seek to characterize the role of neutral hydrogen in and around the galaxies in the reionization epoch \citep[e.g.][]{Jones23,Saxena23,Witstok23} by measuring the Ly$\alpha$ emission line.
The PRISM resolution is particularly coarse at the blue end of the spectrum,
such that the grating substantially outperforms it for the detection of 
narrow lines.  Further, because of the low resolution,
the prism does not distinguish well between mild Ly$\alpha$ emission
and the Ly$\alpha$ damping wing. The medium-deep G140M/F070LP spectra are therefore required
and adequate to measure the Ly$\alpha$ narrow line 
better than the prism and protect the damping wing analyses.

The reductions of these ultradeep spectra were included in the JADES Data Release 3 \citep{DEugenio25}, along with details of the targeting.  
In brief, we have
prioritized high-redshift candidates from JADES, progressively working to
lower redshift.
We included rare special targets
at high priority, incorporating AGN indicators, quiescent galaxies at high redshift, sub-mm detections, and 
strong line emitters extracted from slitless spectroscopy from the FRESCO survey \citep{Oesch23}. As reported in Table \ref{tab:ns}, we could allocate shutters to 228 unique targets. These are slightly fewer targets than allocated in program 1210, because in program 3215 we have given priority in re-observing the same targets in the five pointings, so to maximize the total exposure time for each target.

\begin{figure}[t!]
\includegraphics[width=\columnwidth]{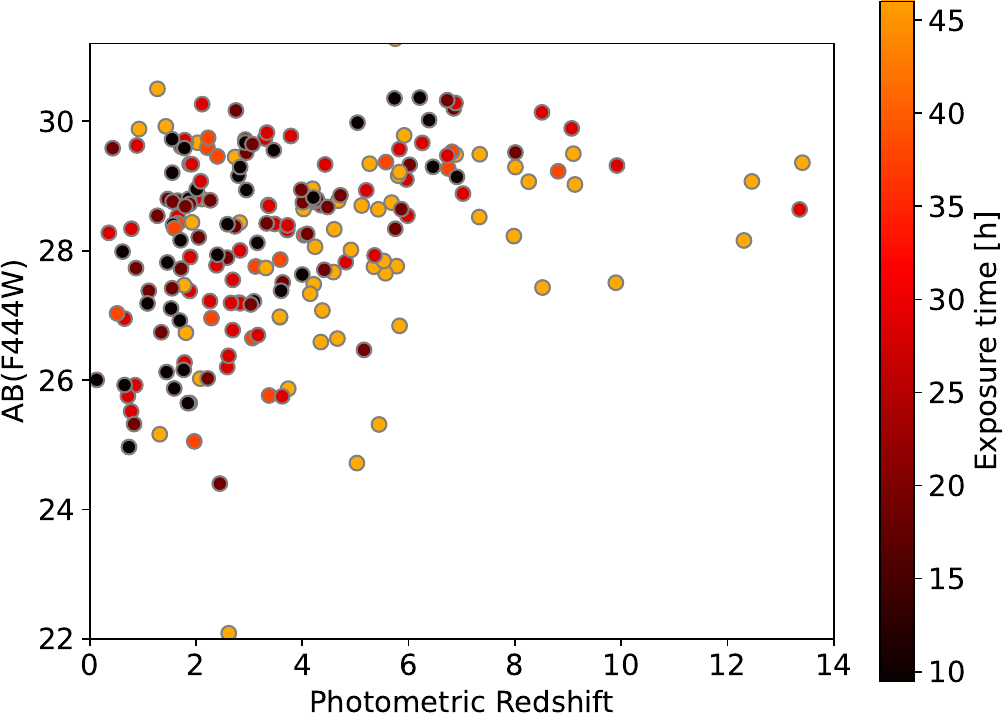}
\caption{\label{fig:ns_targets}%
F444W magnitude versus photometric redshift for the galaxies for which shutters were allocated for NIRSpec observations in program 3215. The color coding indicates the exposure time with the prism for each target.
}
\end{figure}

Figure \ref{fig:ns_targets} shows the distribution of redshift and F444W magnitude for the targets that were allocated shutters in the NIRSpec observation, illustrating that a significant number of them are very faint (${\rm AB}>29$). The color coding gives the total exposure time on each target (for the prism configuration), showing that many targets are observed for the full integration time by being re-allocated shutters in the five pointings.  We note that the three highest redshift galaxies from \citet{Curtis-lake22} are targeted for re-observation.

\subsection{Implementation}

Program 3215 was implemented in the same manner as 1210, described in \citet{Eisenstein23}.  
NIRSpec must be
the prime instrument with NIRCam in coordinated parallel, but by
using multiple MSA configurations at slightly different pointings,
we can achieve a 15-point dither pattern for NIRCam.

We split the program into 5 visits, each with a pair of co-pointed
MSA designs, one for the prism and one for the grating.  Each visit
contains 9 disperser/filter pairs, and each of these is observed
for a 3-shutter nod.  Each exposure contains 2 integrations for NIRSpec
(19 groups of NRSIRS2 readout, \citealt{Rauscher17}, 1400 s each)
as well as 2 integrations for NIRCam (7 groups of DEEP8 readout, 1375 s each).

In each of the 5 visits, we perform 4 nods with the G395M, 1 nod with 
G140M, and finally 4 nods with the PRISM.  These are paired with 6 
NIRCam exposures of F162M \& F335M, 12 exposures of F182M \& F250M, and
9 exposures of F210M \& F300M.  

The exact pointings of the 5 MSA pairs were chosen to place a small number
of $z>11$ candidates onto all of the designs, while providing pixel diversity
for both NIRCam and NIRSpec.  We keep the total spacing small, under 2$''$,
as this increases the number of targets that can be observed on legal 
shutters in most of the designs.  We also use the same position angle
as 1210 (V3 PA of 321$^\circ$) and minimize the displacement of the two
programs, to ensure high overlap of 14 filters.

\subsection{Operational challenges}

The 3215 program was observed October 16--24, 2023, with the data immediately released.  Most of this data is of the hoped-for quality and has been used in numerous papers \citep[e.g.,][]{Robertson24,Hainline24,DEugenio24,Curti25}.  A minor data quality issue is that unsurprisingly we see a return of the blobs and oil slick artifacts that were seen in program 1210, described in \citep{Eisenstein23}.  We mask these in our reductions.  We note that these features also appear (with small translations) in other GOODS-S pointings at this PA, even when the telescope is moved by over an arcminute, but in no pointings we have reduced at other PAs.

\new{A recurring challenge with very deep NIRCam data is the print-through of background residuals from imperfect flat fields.  We encountered this for the wide filters and reported in \citet{Rieke23} that improved sky flats were needed to control the problem.  Unsurprisingly, this also affects the medium-band images, and we have found that pushing to the depth of the JOF requires even better flats.  We note that the dither pattern dictated by the NIRSpec, with 3-step nods and only small transverse steps between visits, appears to exacerbate the meso-scale residuals in the mosaics, even with 15 distinct pointings.  We will discuss this in a future data release paper, showing that the issue is largely resolved at the depth of the JOF.  }

A more serious problem was that the 5th day of observations were impacted by short circuits in the NIRSpec MSA \citep{Rawle22}.  These badly flood the first 18 exposures (6 nods) of this visit, making useless all of the grating exposures and 3 of the 12 PRISM exposures.  The remaining 9 PRISM exposures were not affected.  The total NIRCam exposure times summarized in Table \ref{tab:exptime} do not include these impacted exposures.

This problem also created substantial stray-light banding in long-wave module A of the first 6 NIRCam exposures, all F335M.  The other 12 short-impacted exposures do not show effects, perhaps because the filter changed to F250M, nor is there visible impact in the short-wave data.  This banding must be masked.

The imaging observations also encountered substantial persistence in visits 1, 2, and 4.  In visit 1, the first six exposures, all F162M, have diffuse persistence in modules A3, B3, and B4, typical of a large change in background.  In visits 2 and 4, the A3 data are affected by persistence from prior wavefront sensing, revealing the pupil image and in visit 4 the ingress and egress trails of the target star.  These effects are similar to those encountered in JADES \citep{Eisenstein23}.  As all of the artifacts described above affect only portions of the field, we leave the total exposure times unchanged in Table \ref{tab:exptime}, but note that the effective exposure time on a given location can be mildly shorter.  

We requested and were approved for reobservations of these 6 exposures, both NIRSpec and F162M+F335M imaging, which were planned to occur in October 2024 at the same PA.  Unfortunately, this plan was rendered impossible due to an ill-timed telescope safing.
Facing a delay to October 2025, we developed an alternative plan to instead use the remaining time to conduct spectroscopy on the JOF imaging and to use the NIRCam parallel to create a second shallower field with the same filter set.  This will be described in \S~\ref{sec:second}.

\subsection{High-redshift results from the Origins Field}\label{sec:results}

With the 2023 data, 
\citet{Robertson24} analyzed the $z>12$ UV luminosity function using 9 galaxies in the JOF.  Unfortunately, no $z>15$ galaxies were found, but the upper limits were used to contribute to the inference of the evolution of the UV luminosity function, measuring a drop by a factor of $\sim2.5$ between $z=12$ and 14.  \citet{Adams25} performed a similar analysis in the field of both candidate recovery and the UV luminosity function.

However, the data set did reveal the astonishing JADES-GS-z14-0 \citep{Carniani24}, now spectroscopically confirmed at $z=14.18$ \citep{Carniani25,Schouws25,Schouws25b}.  This galaxy is very bright (45~nJy) and spatially extended, showing that it is not AGN-dominated.  The JOF would have detected it easily a factor of 10 fainter or out beyond $z=20$ \citep{Robertson24}.  Importantly, the medium-band imaging in the JOF was central to selecting the object as a high-redshift candidate; its proximity to a nearby galaxy was confusing until the medium-band data clearly showed a steep dropout at a mildly different wavelength than the Balmer break of the neighbor.  Selection was also encouraged by the deep JADES F770W MIRI imaging, which showed a photometric excess reasonably interpreted as the H$\beta$-[OIII]5007\AA\ emission-line complex \citep{Helton25}.  

With JADES-GS-z14-0 and the recent MoM-z14 \citep{Naidu25}, only twice fainter, it is clear that there is a detectable population of galaxies at $z>15$.  Indeed, the brightness of these two known galaxies might recommend a wider and shallower strategy.  But whether bright or faint, the effects of the dichoic wavelength and the presence of strong emission lines in the LW filters, discussed in \S~\ref{sec:challenges}, should be folded into the design and will likely continue to favor the use of medium bands.

The 3215 imaging data reduction and catalogs will be included in the upcoming JADES data release.

\subsection{Dense-shutter spectroscopy on the JOF}\label{sec:second}

With the opportunity to replan the remaining time into a window in December 2024, we proposed and were approved to change the pointing so as to conduct multi-object spectroscopy on the JOF.  At the same time, the NIRCam parallels at this position angle (V3\_PA 39$^\circ$) fell on deep wide-band imaging from JADES program 1286.  By doing the same medium-band filter set, we could create a second field with the same 15 filters as the JOF, albeit not as deep.  This was conducted on December 17--18, 2024.  

This final pointing consists of a single MSA configuration, which was then observed for 9 nods of a 3-high shutter, providing 3 distinct dither locations.  8 of the nods use long exposures, the same as the original 3215 plan; the 9th is much shorter, only a single integration per position with 6 groups of NIRCam MEDIUM8 readout and 10 groups of NIRSpec NRSIRS2 readout, in order to use the available time allocation.  The NIRCam imaging provides 3 long nods of F250M, F300M, and F210M, 2 long nods of F162M, 2 long and 1 short nod of F182M, and 1 long and 1 short nod of F335M, this last filter being kept short because JADES also provides some of this filter.  We also began the visit with 1 long nod of F150W and F356W, arguing to duplicate this filter from program 1286 as a guard against time-variable sources, which we have found to present false positives for dropout galaxy searches in multi-epoch surveys \citep{DeCoursey25}.

For the spectroscopy, we decided to attempt a more experimental use of the NIRSpec MSA.  JADES had been routinely using designs that create overlapping grating spectra and non-overlapping prism spectra, relying on the prism spectra to reliably sort out which emission lines in the grating data belong to which target.  Here, we opted to omit the prism spectra and observe a single MSA configuration with many overlapping grating spectra, relying on astrometric variations along the slit and priors from the photometric redshifts to associate the lines.  In doing so, we achieve very high multiplex, 850 targets on this one configuration, with a focus solely on emission-line science.  We open shutters on any target that is sufficiently well centered.  The targets were galaxies with photometric redshifts above 3, whose strong emission lines would fall in the G235M and G395M grating, with flux limits down to 7$\sigma$ in the very deep LW imaging.
We omit any galaxy brighter than 300 nJy, as those dispersed continua would be bright enough to add noise to the overlapping extractions.

The approach is motivated by several important technical facts.  First, the NIRSpec grating spectra are detector-noise limited; it requires over 10 overlapping spectra to generate sufficient background to overcome this.  This implies that overlapping the spectra does not generate a harmful amount of photon noise.  Second, the astrometric knowledge from NIRCam gives a strong prior on where lines can occur along the 1.6$''$ length of a 3-shutter slit.  In the wavelength direction, one has additional ways to determine associations from detection of multiple lines and from priors from photometric redshifts and even the detection of the line in the multi-band imaging.  This association problem is familiar from slitless spectroscopy, such as with the NIRCam grism; indeed, it is easier here because one know which targets were given a shutter.  Third, while the density of faint galaxies is high enough that continuum spectra are overlapped, the density of detectable emission lines is low enough that one is very far from the confusion limit.

By using G235M and G395M together on one MSA configuration, we acquire 1.7--5.3 micron spectra, with only small wavelength gaps due to the chip gap.  We observed 4 long nods in each of G235M and G395M, and a short nod in G395M, totaling 33.6 and 35.8 ksec, respectively.  The spectral performance was very good, detecting lines down to 2--$3\times 10^{-19}$ ergs/cm$^2$/s, and we netted 539 successful redshifts in this one MSA configuration.  Further description of the method, along with target selection, resulting performance, and the redshift catalog will be presented in D'Eugenio et al.\ (in prep.).

\section{Cycle 3: NIRCam Grism Spectroscopy with Program 4540}

\subsection{Science Motivation}

While the investment of JWST imaging time in the JOF 
is tremendous, this field does not yet have the deep spectroscopic
attention that the HUDF has received.  

The NIRCam instrument team therefore decided to devote 32 hours of
Cycle 3 guaranteed time to a NIRCam slitless spectroscopy program (program ID 4540, PI: Eisenstein), 
augmenting the MSA coverage from programs 1286, 1287, and 3215.

The NIRCam grism has been shown to produce a compelling capability to
measure strong [OIII] and H$\alpha$ emission lines from galaxies
at $4<z<9$ \citep{Sun23,Kashino23,Matthee23eiger,Oesch23,Wang23}.
By observing in the F322W2, F356W, and F444W filters, program 4540 
produced slitless spectroscopy from 2.4 to 5~$\mu$m over most of the JOF, 
covering the [OIII] or H$\alpha$ lines at $2.7<z<9$.  Of course, rest-infrared lines
are also detected from brighter low-redshift galaxies as well.
Early reductions indicate that these data have measured over 1400 redshifts across 0--9, 
nearly $\sim$100 per square arcminute.

Slitless spectroscopy is an exciting pairing to the JOF deep medium-band imaging, 
as it provides precise redshifts and line shape measurements to the lines inferred from the imaging.
Precise redshifts in turn allow the medium-band photometry to make more accurate inferences
about the continuum breaks, free from uncertainties as to where the weaker lines might fall.
Indeed, the measurement of the line emission in the imaging breaks the degeneracy in slitless spectroscopy between astrometric position and wavelength, allowing the combination to measure 
velocity gradients within a galaxy.
Further, line shapes are a powerful tool to identify Type I AGN and strong galactic outflows
\citep{Matthee23,Maiolino23_bhs,Greene23,Furtak23,Harikane23,Kokorev23}.  

A general census of strong line emitters also provides measures of galactic environment \citep{Kashino23,Witstok23,Helton23,Wang23,Herard-Demanche23,Sun23_850} for studies of SED properties.  Indeed, although this program was originally planned only to include F356W and F444W grism data, the presence of an obvious overdensity of F300M-excess sources in the JOF due to two overdensities at $z=3.47$ and 3.70 encouraged us to split the F356W time to include the F322W2 filter, so that we would have H$\alpha$ coverage of these structures.
In addition, \citet{Helton23} presents the discovery of a substantial arcminute-scale
$z=5.4$ galaxy
overdensity with 43 emission-line galaxy members extracted from the south-west corner of
the GOODS-S FRESCO data \citep{Oesch23}.  Combining with the Cycle-3 pure-parallel NIRCam grism survey SAPPHIRES \citep{Sun2025b,DEugenio2025sap}, Program 4540 has extended this mapping spatially and 
add [OIII] detections to the H$\alpha$ coverage from FRESCO.

Further aspects of the data reduction of this grism data and resulting spectral catalog will be presented in Sun et al.\ (in prep.).

\subsection{Short-wave Data}
We use the opportunity of program 4540 to add deep F070W imaging to the JOF, the final NIRCam broad-band filter and 15th NIRCam filter overall.
Given the mosaic pattern described below, this reaches 
exposure times of 16--68 ks, depending on overlaps, mildly shallower
than but comparable to the 55 and 73 ks in F090W and F115W, respectively.
This extends the JWST coverage of the SEDs and produces optical
imaging much deeper than the ACS imaging from GOODS and CANDELS.

These F070W data provide key enhancements for the core science of
the program, as it provides direct sensitivity to a fainter
population of Ly$\alpha$ break galaxies at $z=5$--6.  This 
yields a UV-selected population to be compared to the emission-line
selected population in [OIII] and H$\alpha$ and to the characterization
of Balmer/4000\AA\ breaks from the medium-band data.

\subsection{Implementation}

\begin{figure*}[!t]
\centering
\includegraphics[width=\linewidth]{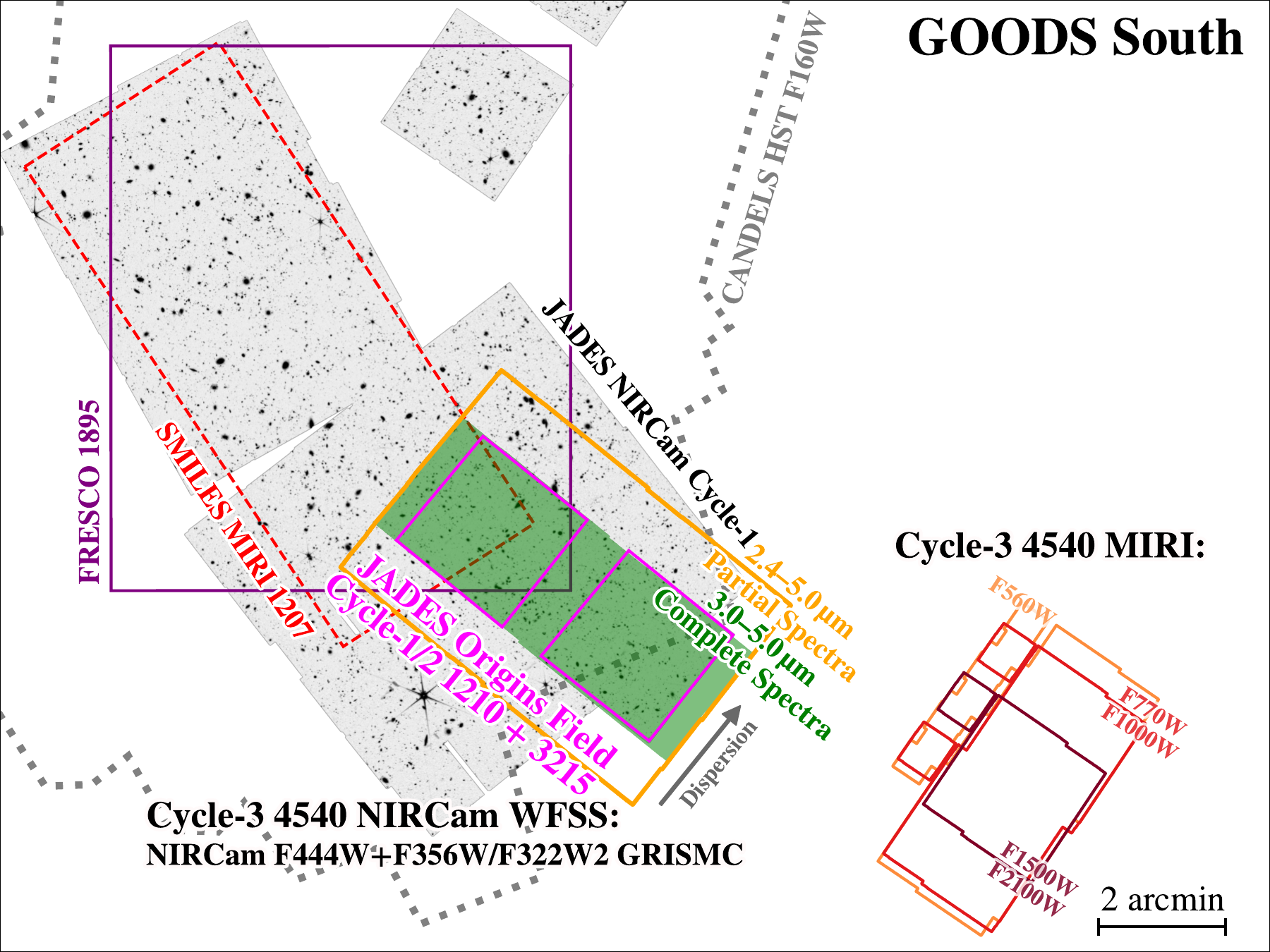}
\caption{Layout of JADES Cycle 3 observations in the GOODS-South field.
Cycle 1 JADES NIRCam imaging observations are shown as the background image. 
The JOF regions are shown with solid magenta line.
The spectral coverage region of the Cycle 3 NIRCam WFSS GRISMC observations are shown as green shaded regions (for complete spectral coverage at 3--5\,\micron) and solid orange lines (for partial spectral coverage across 2.4--5\,\micron), respectively.
The dispersion direction of GRISMC is indicated as the gray arrow.
To the west of the NIRCam footprints, the coordinated MIRI imaging parallel observations are indicated by the solid orange (F560W), red (F770W and F1000W) and maroon lines (F1500W and F2100W), respectively.
For comparison we also plot the footprints of CANDELS HST F160W imaging \citep[dotted grey lines;][]{Whitaker19}, SMILES MIRI multi-wavelength imaging (Cycle-1 GTO 1207; \citealt{alberts2024}, \citealt{riekeg2024}), and FRESCO NIRCam imaging and F444W GRISMR WFSS \citep{Oesch23}.
}
\label{fig:cy2cy3}
\end{figure*}

Driven by the desire to cover the 3 to 5~$\mu$m range over the 
full program 1210 footprint as well as to cover the intermodule gap and a small
boundary for environment measurements, we designed program 4540 for a tight 2$\times$2
mosaic. %
As shown in Figure~\ref{fig:cy2cy3}, the same data give partial wavelength coverage across 2.4--5.0~$\mu$m in a mildly broader area;
this bounding area is largely covered by JADES medium-depth NIRCam imaging.

As redshifts in this field are so valuable, we aim to achieve a
depth that is notably fainter than the FRESCO observations ($5\sigma$ depth of unresolved emission line is $\sim2\times10^{-18}$\,erg\,s$^{-1}$\,cm$^{-2}$ around 4.3\,\micron).  In the overlap regions
of the 4 tight pointings, we reach a maximum depth of $\sim10.9$\,hrs of integration
time in the F444W filter, compared to $\sim$2 hrs in FRESCO F444W.  We had originally planned
for an equal amount of time in F356W, but later decided to split those exposures evenly between the F356W and F322W2 filters, noting that the
F356W filter data is yet more sensitive because of lower zodiacal
background.   
Preliminary analyses suggest that the data have reached a $5\sigma$ depth for unresolved emission lines of $\sim0.9\times10^{-18}$\,erg\,s$^{-1}$\,cm$^{-2}$ around 4.3\,\micron\ (with the F444W filter at a median exposure time of 5.4\,hrs); $\sim1.0\times10^{-18}$\,erg\,s$^{-1}$\,cm$^{-2}$ around 3.6\,\micron\ (with the F356W filter alone at a median exposure time of 3.0\,hrs); and $\sim3.5\times10^{-18}$\,erg\,s$^{-1}$\,cm$^{-2}$ around 2.8\,\micron\ (with the F322W2 filter at a median exposure time of 3.0\,hrs).

The grism data are undeflected at 3.95\,\mic, so the regions of complete F356W spectra and complete F444W spectra fall in different portions of the NIRCam field of view.  
With the column-direction grism (GRISMC), we use 2 rows of the mosaic stepped so that we can get complete coverage over 
the full JOF NIRCam imaging footprint.  One observation observed in all 3 filters; this produces coverage maps that are displaced as a function of wavelength.
The other 2 observations were offset, with F444W toward the northwest and F356W/F322W2 toward the southeast, so that we get complete 3 to 5~\mic\ 
coverage over the JOF.  
GRISMC is preferred to GRISMR for this purpose because the dispersion direction is the same in the two modules.
Blueward of 3.1 microns, the F322W2 data (which use the same pointing as F356W) sample an even smaller field, with gaps opening in the footprint at the very blue end.

We use 6 dither pointings in each of the mosaic pointings, using the INTRAMODULEBOX pattern.  Because of allocation constraints, we have different exposure times.  Two of the pointings use DEEP8 readout, with 5 groups for F322W2 and F356W and 8 groups for F444W.  The other two pointings use 8 groups of MEDIUM8 readout, with F444W being observed twice.

Although slitless spectroscopy often uses multiple dispersion directions 
to robustly connect lines to the zero-order location and hence get unique wavelengths,
we do not do this here.  With the sharp NIRCam imaging and accurate grism spectral tracing models, the associations of lines to galaxies detected in the imaging is unconfused \new{in most cases}.
More important, because of the deep multi-band imaging,  
the high-equivalent emission lines (essentially all of the $z>4$ cases) 
show up easily in the medium-bands
from programs 1180, 1210, and 3215.  This sharply reduces concerns about
the ability to robustly match up the line detections to the undispersed images.
\new{Combining with the emission-line maps from imaging can even break the degeneracy between morphology and velocity fields inherent to slitless spectroscopy \citep[e.g.,][]{Nelson23,danhaive25}.}
Further, the NIRCam imaging yields very high quality photometric redshifts,
which reduces the pressure to have a second line detection to produce an identification of the line and hence a redshift.
\new{That said, the lack of multiple dispersion angles does sharply limit the ability to robustly extract continuum spectra from the grism data.}

Direct images are required by the slitless spectroscopy observing template.
We use short exposures (6 groups of SHALLOW4, 311 sec) because this 
field already has far deeper data.  We use the same LW 
filters, but change in SW to F150W and F200W to increase the signal-to-noise in the 
short exposures, to assure some infrared coverage in the southwest edge of field where we may fall in gaps of other NIRCam imaging, and to get a 3rd separate year of data for time-domain studies.

\subsection{Coordinated Parallels with MIRI}

Program 4540 also produced coordinated parallels for MIRI imaging in five filters.  
The longer grism observation with all 3 LW filters is used for F1500W and F2100W; this produces the tightest footprint about 2$'$ square with total depth of about 7 ks per filter . 
The other two observations use F770W and F1000W, producing a $4'$$\times$$2'$ footprint centered on the first with total depth of about 5 ks per filter.
\new{We note that one of the four MIRI F1000W pointings was lost because of insufficient disk space on the spacecraft, resulting in reduced coverage and depth at the southeastern corner of the mosaics. }
All of the direct imaging use F560W, producing a mildly larger footprint; the increased sensitivity of F560W will
compensate for the shallower depth of 1--2 ks.
The footprints of coordinated MIRI parallels are also shown in Figure~\ref{fig:cy2cy3}.
The longer observation uses FASTR1 readout; the other two use SLOWR1 to limit the data volume.
Aside from F560W, we use multiple integrations per NIRCam integration to mitigate saturation.
In detail, the number of integrations and groups are 3 \& 11, 3 \& 11, 3 \& 71, and 14 \& 31 in F770W, F1000W, F1500W, and F2100W, respectively; F560W uses either 13 slow or 100 fast groups.
The JWST ETC forecasts 5-sigma depths of 25.0-25.3, 25.5, 25.0, 24.5, and 22.9 AB mag in the F560W, F770W, F1000W, F1500W, and F2100W bands, respectively, assuming a point source with an 0.6$''$ diameter aperture.

These F770W and F1000W data will extend the study of $z>4$ galaxies and
particularly their stellar mass utilizing rest-frame 1--2~\mic\
measurements, while the longer wavelength data will provide insight into the role
of AGN in $z\approx 2$--5 galaxies, as the hot dust in the nuclei
can outshine the Rayleigh-Jeans tail of the stellar population.
Finding such sources may be the most powerful means to identify heavily obscured AGN at $z>5$.

However, the location of this parallel field does fall into an area 
currently lacking in NIRCam and HST imaging.  Of course, the extended
CDF-S does have deep ground-based imaging (such as the deep 5-year stack from the Dark Energy Survey SNe program and the recently released Rubin DP2) as well as imaging from Spitzer,
which can support near-term use.   We expect that this field is likely
to be imaged with NIRCam in the future, as its location
is easy to reach as a parallel when NIRSpec is pointed at the 
exceptional JOF NIRCam imaging, indeed at the position angle
such that the NIRSpec quadrants align to the NIRCam geometry.

\section{JADES Second Imaging Data Release}

\begin{figure*}[p]
\includegraphics[width=\linewidth]{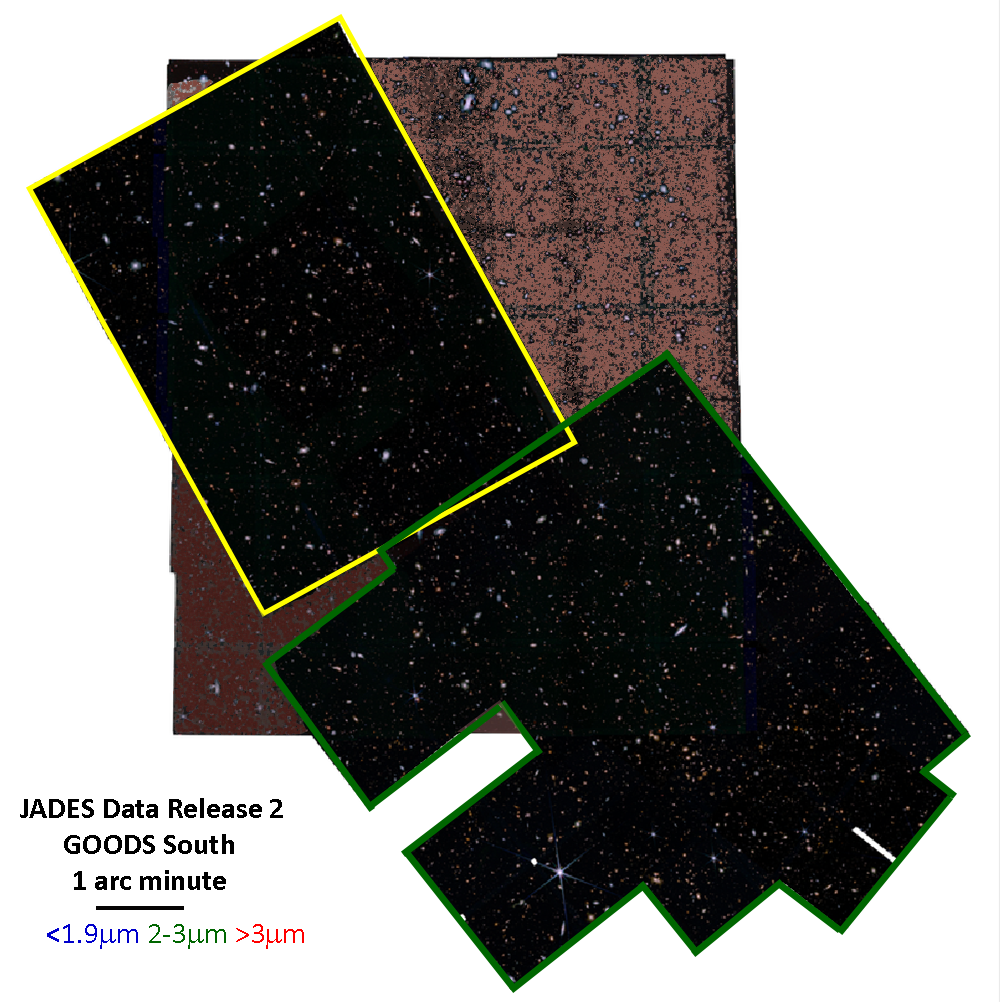}
\caption{\label{fig:fieldimage}%
A rendering of the JADES GOODS-S data release image, combining filters 0.9--1.9~\mic\ in blue, 2--3~\mic\ in green, and 3--5~\mic\ in red.  The green boundary indicates the region new to this release; the yellow boundary indicates the region also in the first JADES release \citep{Rieke23}.  The region where the imaging is solely from the FRESCO survey \citep{Oesch23} is tinted brown to distinguish it.
}
\end{figure*}

\begin{figure*}[p]
\includegraphics[width=\linewidth]{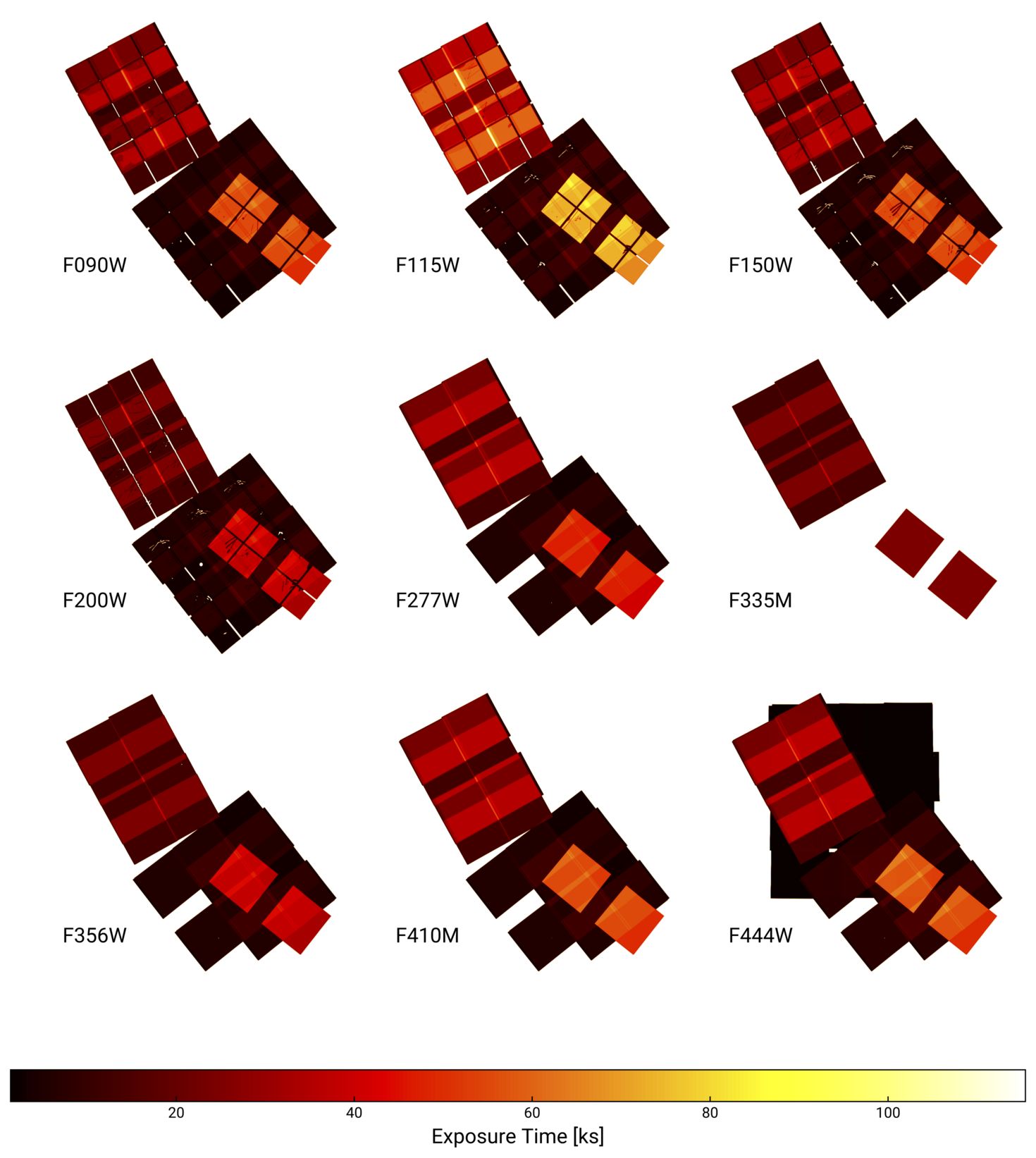}
\caption{\label{fig:depth}Exposure maps of the JADES NIRCam F090W, F115W, F150W, F200W, F277W, F335M, F356W, F410M, and F444W imaging. The 1180 Deep program comprises the rectangle in the northeast and 1180 Medium the wider, shallower rectangular exposure area to the southwest. The JADES 1210 imaging of the JOF appears as a deep module pair in the southwest. The F335M filter is only used in 1180 Deep and 1210. The F444W image includes the wide-area FRESCO imaging \citep{Oesch23}. Not shown are the F182M and F210M exposure maps from the JEMS \citep{Williams2023} and FRESCO program data, or the F430M, F460M, and F480M JEMS imaging.}
\end{figure*}

\begin{figure*}[t]
\noindent\includegraphics[width=0.3\linewidth]{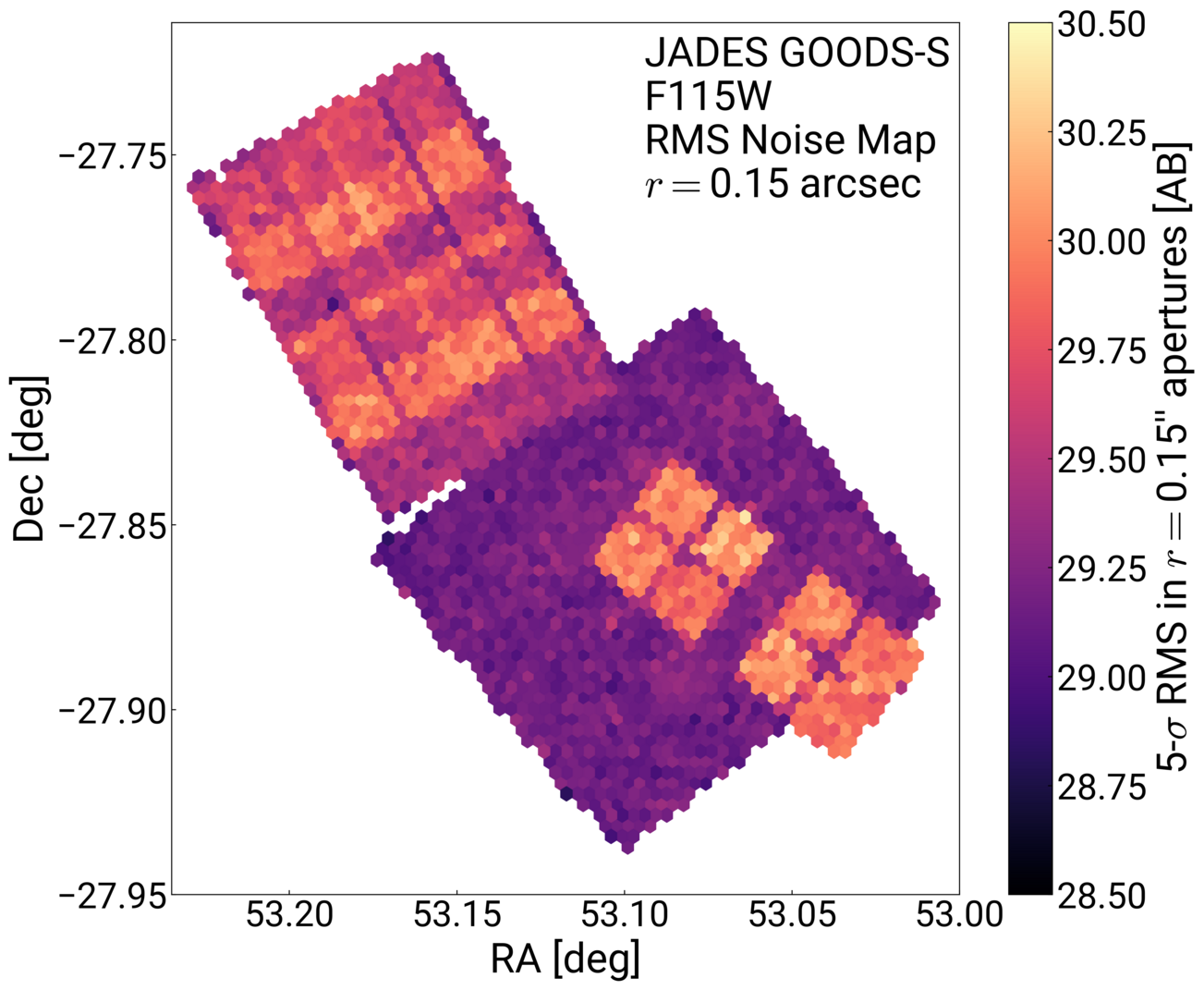}\hfill
\includegraphics[width=0.3\linewidth]{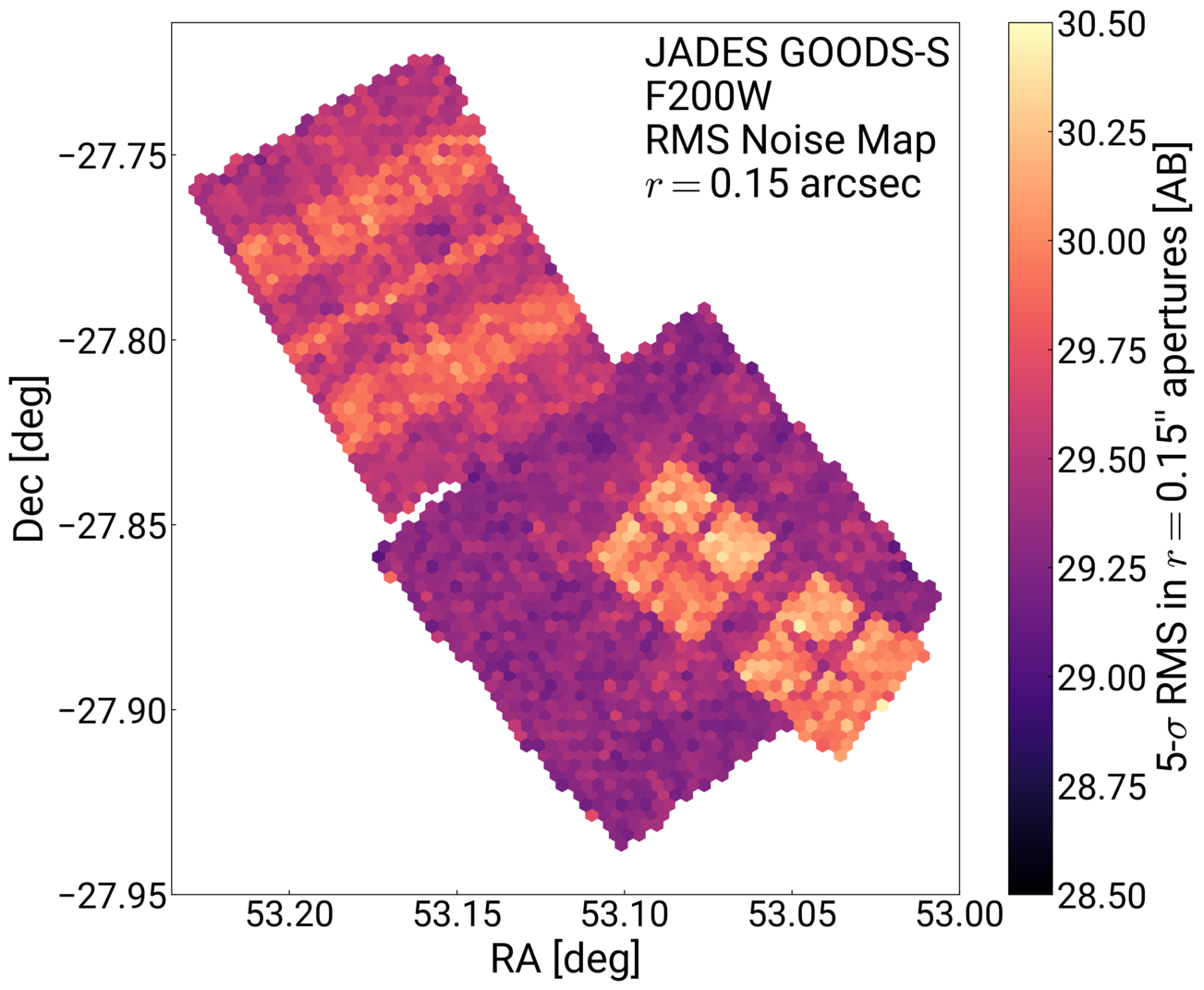}\hfill
\includegraphics[width=0.3\linewidth]{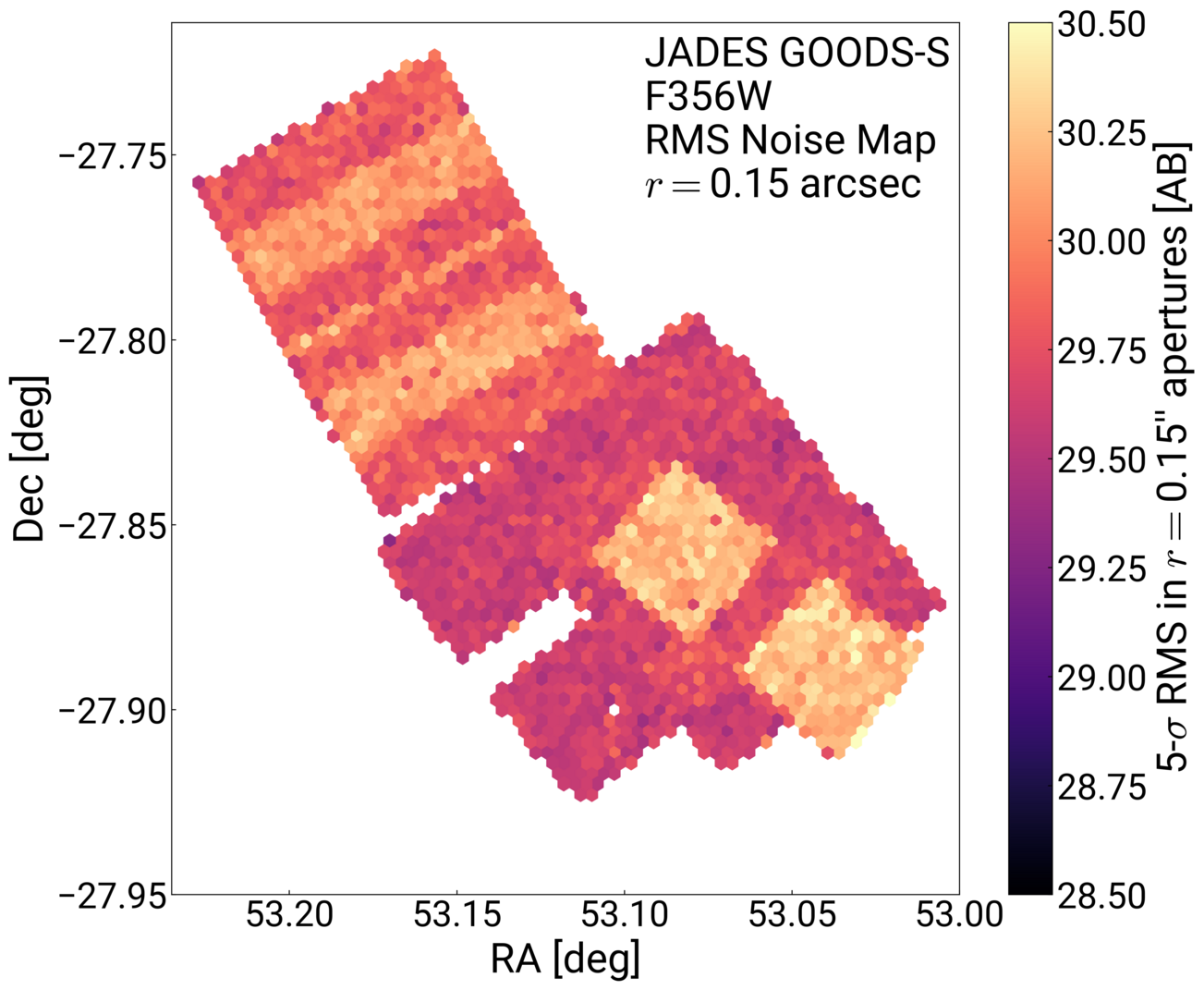}\\
\caption{\label{fig:magdepth}The depth in the F115W, F200W, and F356W bands, expressed as the 5$\sigma$ AB magnitude for an $0.15''$ radius aperture, measured from the rms scatter from empty regions of the mosaic and corrected for the point source curve of growth.  We quote this aperture for comparison with \citet{Rieke23} and previous HST work, but we note that an $0.1''$ aperture yields yet deeper limits, by about 0.3 mag in F200W.}
\end{figure*}

In this paper, we present the second imaging data release of JADES, including the GOODS-S imaging observed in September and October 2022.  The first release \citep{Rieke23} included the NIRCam imaging of the Deep Prime program on the HUDF.  This second release includes the deep NIRCam parallel from program 1210 and the medium-depth mosaic that flanks these two imaging fields.  Caveats regarding data quality and some missing data due to short circuits in the NIRSpec MSA and one skipped visit are described in \citet{Eisenstein23}.

Figure \ref{fig:fieldimage} shows a three-color image of the field.  The total area of the JADES imaging (excluding the FRESCO-only region) is 67.7 arcmin$^2$, and the catalogs contain 94,000 distinct objects. A summary of the image reduction and catalog processing follows.

To create calibrated imaging and object photometry catalogs, the NIRCam data are reduced in the same manner as described in \citet[][hereafter R23]{Rieke23} with a few modifications. For this release, version 1.11.4 of the STScI \emph{jwst} pipeline
is used but with a custom astrometric system as discussed in R23. The JWST calibration reference data file used during processing was {\tt jwst\_1130.pmap}, which adjusted the photometric zeropoints slightly relative to the first JADES data release. We have checked that the flux ratios of objects shared between the current data release and \cite{Rieke23} are broadly consistent with the changes in photometric zeropoint.

To increase our pixel redundancy, the JADES V2.0 mosaics are constructed by simultaneously stacking all available exposures at each location rather than the visit-based combination and stacking used in R23. To achieve this without taxing computer memory, we split each mosaic into smaller sub regions that can be processed separately. For F090W, F115W, and F150W, we remove a small number of hot pixel residuals from the pipeline by replacing the affected pixels with the corresponding pixels from median-filtered versions of the mosaics. After visual inspection of individual exposures, we masked regions of high persistence or with stray light effects with high spatial frequency, especially noticeable in the 1180 Medium region. This masking reduces the depth and exposure time in only a small fraction of the total JADES area.

The exposure times achieved in this mosaic are shown in Figure \ref{fig:depth}.  These show the deep parallel field of program 1210, along with the year 1 data in the Deep and Medium mosaics of program 1180.  
Year 2 observations doubled the depth of the Deep mosaic, added flanking Medium depth around the Deep region, and extended the footprint with parallel NIRCam imaging from program 1286 and 1287 \citep{Eisenstein23}.
Figure \ref{fig:magdepth} shows maps of the corresponding 5$\sigma$ sensitivity for F115W, F200W, and F356W in $0.3''$ diameter apertures, with aperture corrections applied. These maps are computed by placing randomized apertures throughout the fully reduced mosaic images, after masking sources and surrounding pixels with significant flux.
\new{Such errors are preferred to formal errors derived from pixel-level uncertainties because of the correlations between pixels induced by the mosaicking.}
As reported by \citet{Rieke23}, the 1180 Deep region reaches F200W depth of $29.8$ mag AB. 

To enhance the multi-band space-based photometry available for sources, we also perform photometry on the Hubble Legacy Field HST/ACS F435W, F606W, F775W, and F814W mosaics and HST/WFC3 F105W, F125W, F140W, and F160W mosaics \citep{illingworth16,Whitaker19} after rectifying them to our astrometric frame. We reprocess and release JWST/NIRCam F182M and F210M mosaics from data acquired by the JWST Extragalactic Medium-band Survey \citep[JEMS;][]{Williams2023} and First Reionization Spectroscopically Complete Observations \citep[FRESCO;][]{Oesch23} programs.
We additionally reprocess and release JWST/NIRCam F430M, F460M, and F480M images from data acquired by JEMS and combine reprocessed JWST/NIRCam F444W exposures originally acquired by FRESCO with our F444W stack. We include photometry from all JWST filters in our released catalogs, where available.

The construction of the photometric catalogs largely
follows that described by \citet{Rieke23}. The LW channel images are stacked to create an
inverse-variance-weighted signal-to-noise ratio image for performing detection. Catalogs are constructed 
from the resulting segmentations, with automated and manual masking of stars and diffraction spikes applied.
SExtractor \citep{Bertin1996}
windowed positions are used as object centroids. Aperture photometry
is measured using a range of circular apertures, two Kron apertures, and segmentations as discussed in \citet{Rieke23}. Sky background uncertainties are computed by placing random apertures throughout the images and directly from the \emph{jwst} pipeline ERR image layers when available. Aperture corrections are computed from model point-spread-functions constructed following \citet{Ji23}. Flags are set to indicate
bad pixels that may impact object segments, whether
sources may be influenced by brighter neighbors,
and whether stars identified in the image may impact them or if the objects themselves are consistent with 
stars.

In addition to the photometry, we have computed photometric redshifts following the methods described in \citet{Rieke23} and \citet{Hainline23}.  In particular, these use the EAZY code \citep{Brammer08} with templates as described in \citet{Hainline23} and incorporating percent-level adjustments to the photometric zeropoints  based on the global median offset of each filter in a first-round of fitting.
For convenience, the photometric redshift catalog is included as FITS extension of the photometry catalog.  \new{A comparison to spectroscopic calibrators was plotted for DR1 in \citet{Rieke23} and \citet{Hainline23}, showing good performance in the aggregate.  Of course, calibrations for faint populations are meager, and one should be careful about how outliers, particularly of unusual colors, might impact one's results.  We note that with the medium-band photometry, galaxies with strong rest-optical emission lines often have more robust photometric redshifts, so performance is not a simple function of continuum magnitude.}

The images and catalogs are available on the Mikulski Archive for Space Telescopes (MAST) at \url{https://archive.stsci.edu/hlsp/jades}, with DOI 10.17909/z2gw-mk31.
We will incorporate the 3215 imaging using this same pipeline in an upcoming release.

\begin{figure*}[t]
\includegraphics[width=\linewidth]{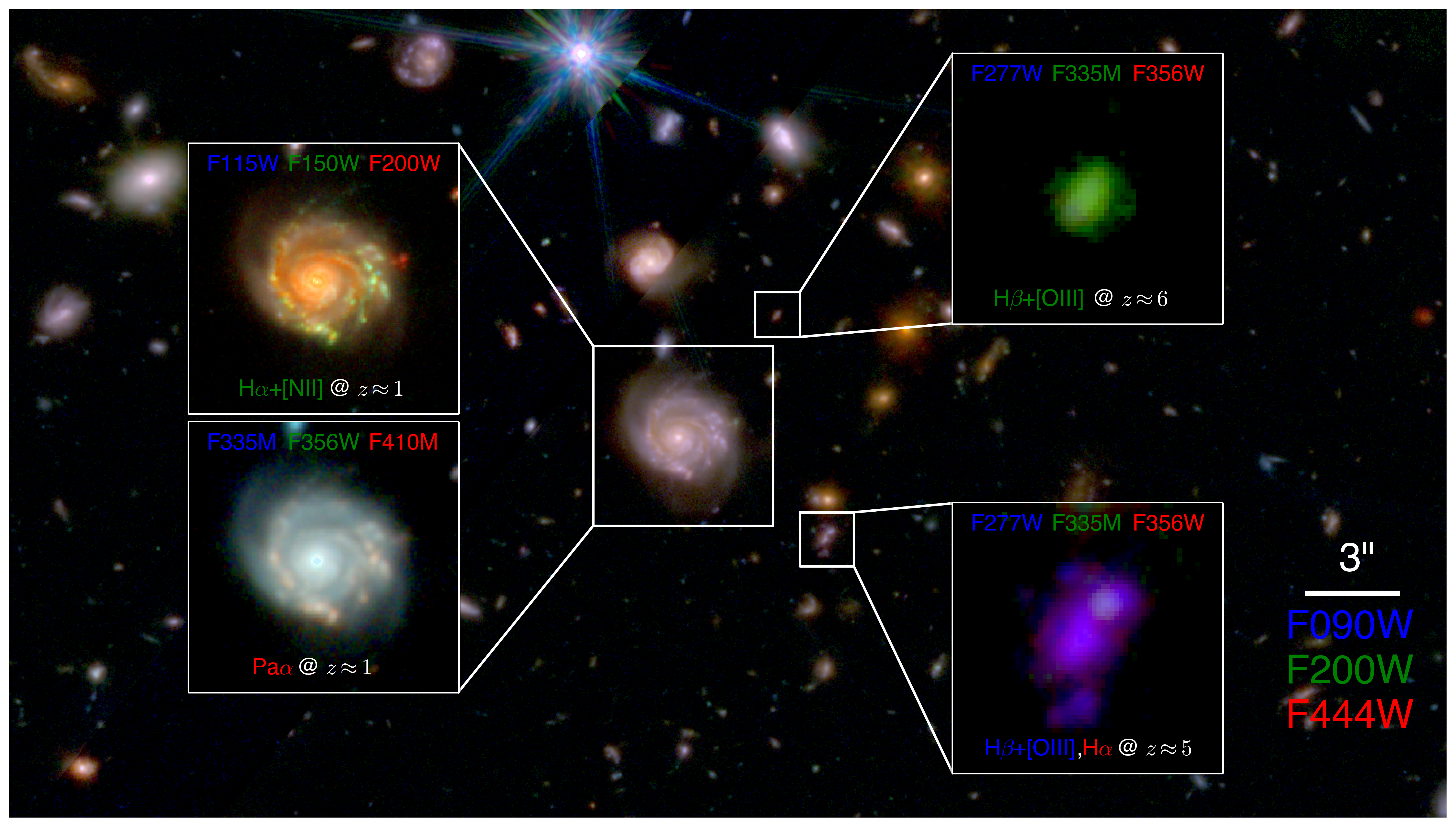}
\caption{\label{fig:image}%
A small portion of the deep imaging from the JOF, illustrating some of the opportunities of the medium-band data. The background mosaic shows an F444W/F200W/F090W RGB image with three galaxies highlighted. A disk galaxy at redshift $z\sim1$ (center) shows star forming regions that glow distinctively in various JADES filter combinations. Emission from H$\alpha$ and [NII] appear as green in an F200W/F150W/F115W RGB image, and Pa$\alpha$ appears as red in an F410M/F356W/F335M RGB image. The redshifts of line-emitting galaxies can be easily identified using the JADES medium bands. At $z\sim6$, H$\beta$+[OIII] appear as green in an F356W/F335M/F277W RGB image. At z$\sim5$, the presence of H$\beta$+[OIII] in F277W, H$\alpha$ in F356W, and no strong line in F335M will cause a line-emitting galaxy to appear purple in an F277W/F335M/F356W RGB image.  \new{As is common in JWST mosaics of data taken an multiple position angles, chip gaps in some of the input images produce disjoint diffraction spikes in the bright star at the top.}
}
\end{figure*}

\section{Discussion}

We have described the JWST observations
of the JADES Origins Field, spanning the first three mission cycles.
In total, the time invested in this single NIRCam footprint of
$\sim$9 arcmin$^2$ is already very large, roughly 360 open-shutter
hours of exposure time across in 15 NIRCam imaging filters, 20 hours
of NIRCam slitless spectroscopy, and over 170 hours of MIRI imaging.
This is the most intensively observed JWST field yet, designed to probe
the redshift frontier beyond $z=15$.

Figure \ref{fig:image} shows a small portion of the JOF imaging data from 
program 1210 to highlight the opportunities of the deep NIRCam medium-band data, 
here F335M and F410M.
Strong emission lines are revealed easily in these images, with beautiful morphological
detail.  This reach was greatly extended in program 3215 with 5 new medium bands
spanning 1.6 to 3.0 \mic.

The location of the JOF in close proximity to the HUDF and the
Deep Prime JADES NIRCam imaging mosaic adds considerable value to the
GOODS-S region.  Wider field telescopes can capitalize on both regions,
and the parallel opportunities will continue to appear with JWST.
By construction, further NIRSpec observations of the HUDF can place further
NIRCam exposure time in the JOF region; the position angle is at a
period of low zodiacal background and is allowable with the micrometeorite 
constraint.  Indeed, program 3215 used this opportunity to conduct exceptionally deep spectroscopy (91 hrs of total exposure) in the HUDF, focusing on the physics of high-redshift galaxies.  Further, observations with NIRISS or MIRI on the JOF are
easily paired with NIRCam observations on and around the HUDF, within the JADES Deep NIRCam mosaic.

Accompanying this paper, we have released the reduced images and catalogs of the Year 1 JADES data in and around the JOF.  These data are already among the very deepest observed with JWST NIRCam, with over 110 hrs of total exposure spread across 9 filters.  These data are available at \url{https://archive.stsci.edu/hlsp/jades} and a feature-rich FITSmap viewer is provided at \url{https://jades.idies.jhu.edu}, with a version tailored for the public at 
\url{https://jades.idies.jhu.edu/public}.

The power of JWST to reveal the early history of galaxies with infrared imaging and spectroscopy is both transformative and inspiring.  Many compelling observing programs have been performed or are underway, each with their own nuances and opportunities.  We hope that the extraordinary depth and filter complement of the JADES Origins Field provides a superb new view of galaxies in their youth.

\bigskip

\begin{acknowledgements}
The JADES Collaboration thanks the Instrument Development Teams and the instrument teams at the European Space Agency and the Space Telescope Science Institute for the support that made this program possible. We also thank our program coordinators at STScI for their help in planning complicated parallel observations.

This work is based in part on observations made with the NASA/ESA/CSA James Webb Space Telescope. The data were obtained from the Mikulski Archive for Space Telescopes at the Space Telescope Science Institute, which is operated by the Association of Universities for Research in Astronomy, Inc., under NASA contract NAS 5-03127 for JWST. These observations are associated with programs 1180, 1210, 1895, 1963, 3215, and 4540.  We acknowledge the FRESCO and JEMS teams for developing their observing programs with a zero-exclusive-access period.

Processing for the JADES NIRCam data release was performed on the \emph{lux} cluster
at the University of California, Santa Cruz, funded by NSF MRI grant AST 1828315.
This research makes use of ESA Datalabs (datalabs.esa.int), an initiative by ESA’s Data Science and Archives Division in the Science and Operations Department, Directorate of Science.
This work was performed using resources provided by the Cambridge Service for Data Driven Discovery (CSD3) operated by the University of Cambridge Research Computing Service (www.csd3.cam.ac.uk), provided by Dell EMC and Intel using Tier-2 funding from the Engineering and Physical Sciences Research Council (capital grant EP/T022159/1), and DiRAC funding from the Science and Technology Facilities Council (www.dirac.ac.uk).

Support for JEMS program JWST-GO-1963 was provided in part by
NASA through a grant from the Space Telescope Science Institute,
which is operated by the Associations of Universities for Research
in Astronomy, Incorporated, under NASA contract NAS 5-26555.

MR, PC, EE, DJE, JMH, BDJ, BR, GR, FS, and CNAW acknowledge support from the NIRCam Science Team contract to the University of Arizona, NAS5-02015.  We also have been supported by program JWST-GO-3215.  DJE is further supported as a Simons Investigator.  
\end{acknowledgements}

\begin{acknowledgements}
SAr acknowledges support from Grant PID2021-127718NB-I00 funded by the Spanish Ministry of Science and Innovation/State Agency of Research (MICIN/AEI/ 10.13039/501100011033).  
AJB, AJC, JC, AS \& GCJ acknowledge funding from the ``FirstGalaxies" Advanced Grant from the European Research Council (ERC) under the European Union’s Horizon 2020 research and innovation programme (Grant agreement No. 789056).    
ECL acknowledges support of an STFC Webb Fellowship (ST/W001438/1).  
Funding for this research was provided to RH by the Johns Hopkins University, Institute for Data Intensive Engineering and Science (IDIES).  The Cosmic Dawn Center (DAWN) is funded by the Danish National Research Foundation under grant no.140.  
RM, WB, FDE, TJL, JS, LS, and JW acknowledge support by the Science and Technology Facilities Council (STFC) and by the ERC through Advanced Grant 695671 ``QUENCH" and the UKRI Frontier Research grant RISEandFALL. RM also acknowledges funding from a research professorship from the Royal Society.  JW further acknowledges support from the Fondation MERAC.  The research of CCW is supported by NOIRLab, which is managed by the Association of Universities for Research in Astronomy (AURA) under a cooperative agreement with the National Science Foundation.  
RS acknowledges support from a STFC Ernest Rutherford Fellowship (ST/S004831/1).  CWo is supported by the National Science Foundation through the Graduate Research Fellowship Program funded by Grant Award No. DGE-1746060.  DP acknowledges support by the Huo Family Foundation through a P.C.\ Ho PhD Studentship.  H{\"U} gratefully acknowledges support by the Isaac Newton Trust and by the Kavli Foundation through a Newton-Kavli Junior Fellowship, as well as funding by the European Union (ERC APEX, 101164796). Views and opinions expressed are however those of the authors only and do not necessarily reflect those of the European Union or the European Research Council Executive Agency. Neither the European Union nor the granting authority can be held responsible for them.
SC and GV acknowledge support by European Union’s HE ERC Starting Grant No. 101040227 -- WINGS.

\end{acknowledgements}

\software{
Aladin Sky Atlas \url{http://aladin.u-strasbg.fr/};
Astronomer’s Proposal Tools \url{https://www.stsci.edu/scientific-community/software/astronomers-proposal-tool-apt};
eMPT code \protect\citep{bonaventura23a};
dynesty \protect\citep{Speagle20};
FitsMap \protect\citep{hausen22};
grizli \url{https://doi.org/10.5281/zenodo.7963066};
Montage \url{http://www.ascl.net/1010.036};
NCDhas (Misselt, private communication);
Prospector \protect\citep{Johnson21};
SExtractor \protect\citep{Bertin1996}
}

\clearpage
\bibliography{biblio}

\end{document}